\documentclass[draft]{agujournal2019}
\usepackage{url}
\usepackage{lineno}
\usepackage[inline]{trackchanges} 
\usepackage{array, mathtools}
\usepackage{amsmath,amsfonts,amssymb}
\usepackage{soul}
\usepackage{bm}
\usepackage{longtable}
\usepackage{booktabs}
\usepackage{caption}
\begin{document}
	\justifying
%	\linenumbers
	\title{The role of thermal buoyancy in
		stabilizing the axial dipole field in 
		rotating two-component convective dynamos}
	
	\authors{Debarshi Majumder and Binod Sreenivasan\thanks{bsreeni@iisc.ac.in (Binod Sreenivasan)}}
	\affiliation{}{Centre for Earth Sciences, 
		Indian Institute of Science, Bangalore 560012, India}
	
	%	\correspondingauthor{Binod Sreenivasan}{bsreeni@iisc.ac.in}
%\vspace*{-0.25cm}
%\date
\begin{center}
28 March 2026
\end{center}

%\vspace*{-2cm}
	
%	\begin{keypoints}
		%%
%		\item The range of compositional buoyancy that
%		admits the dipole in a two-component dynamo is much wider 
%		than that in the
%		one-component dynamo.
		
%		\item Compositional buoyancy $\sim 10^3$ times 
%		its value for convective onset and thermal buoyancy 
%		near above onset produce the observed polar vortices.
		
%		\item Since a thermochemical dynamo 
%		resides deep in the dipolar regime, 
%		a large heat flux variation on Earth's CMB induces
%		polarity reversals.
		%%
%	\end{keypoints}

	\begin{abstract}
		\justifying
		Two-component convection driven
		by both compositional and thermal buoyancy
		within the fluid core of a rapidly
		rotating planet produces a predominantly axial dipole
		field. In a dynamo driven by
		strong compositional buoyancy
		that by itself destabilizes the axial dipole, 
		the addition of relatively weak thermal buoyancy
		establishes the dipole field through the 
		spontaneous generation of slow magnetostrophic
		waves produced by balances between the magnetic,
		buoyancy and Coriolis (MAC) forces at several
		locations within the core. 
		A substantially higher compositional
		buoyancy is then required to trigger polarity transitions, 
		since the dipolar 
		regime is extended 
		in two-component convection, as predicted by
		a linear magnetoconvection model that analyses the long-time
		evolution of a density disturbance.
		The 
		existence of the axial dipole also prescribes 
		a lower bound
		for the fraction of the total power
		contributed by thermal buoyancy, $\approx 10$\%, above
		which the two-component dynamo with homogeneous
		boundary heat flux lies deep within the dipolar regime.
		Two-component convection has
		implications for Earth's core dynamo: dominant compositional
		buoyancy ensures the observed polar circulation
		speed, 
		and a large heterogeneity in the lower-mantle heat flux
		induces magnetic field excursions and
		occasional polarity reversals.
		
	\end{abstract}

	\section{Introduction}
	Earth's dipole-dominated magnetic
	field is generated 
	by a convective dynamo operating within its
	outer core. Following the
	formation of Earth's inner core,
	thought to have occurred less than
	a billion years ago \cite{labrosse2001age,bono2019young}, compositional
	buoyancy arising from the continual release of
	light elements at the inner core boundary  \cite{poirier1994light}
	has driven turbulent motion in the
	core. Thermal buoyancy, on the
	other hand, arises from secular cooling of the
	planet throughout its history
	as well as 
	the release of latent heat during 
	inner core solidification. 
	Early Earth, modelled
	without a solid inner core, was driven
	solely by thermal buoyancy \cite{lin2025invariance}.
	In present-day 
	Earth, the primary source of 
	buoyancy for outer core convection 
	is thought to be compositional,
	contributing up to 80\% of 
	the total convective power 
	\cite{lister1995strength}.

	In numerical dynamo models, 
	the two species of
	temperature and composition 
	are often merged into a single \lq codensity' 
	variable, under the assumption 
	of comparable turbulent 
	diffusivities 
	\cite{braginsky1995equations}. 
	%This approach reduces the 
	%two transport equations to 
	%one and significantly 
	%lowers computational cost. 
	In reality, the ratio of thermal diffusivity 
	to compositional diffusivity, $\kappa^T/\kappa^C$ 
	is typically of 
	$O(10^3)$ \cite[p.~298]{cormier2021earth},
	which gives a thermal Prandtl number 
	$Pr = \nu/\kappa^T = O(10^{-1})$ and a compositional 
	Schmidt number $Sc = \nu/\kappa^C = O(10^2)$, where 
	the kinematic viscosity 
	(or momentum diffusivity) $\nu \sim O(10^{-6})$ 
	\cite{pozzo2012thermal}. That said, $\nu$,
	$\kappa^T$ and $\kappa^C$ are all
	small in comparison with the magnetic diffusivity $\eta$,
	the finiteness of 
	which places bounds on the length scale
	of a buoyancy disturbance, $\delta$ 
	that can be supported against magnetic
	diffusion as well as the square of the peak field
	intensity, $B_{peak}^2$ \cite{sreenivasan2017confinement,jfm21}. Therefore, regardless of the
	ratio $\kappa^T/\kappa^C$, we anticipate that
	the evolution of isolated
	buoyancy disturbances in a two-component convective
	dynamo would be well approximated by 
	linear magnetoconvection
	in the limit $\nu , \kappa^C, \kappa^T \ll \eta $, which leads to
	
	\begin{equation}
		\frac{\nu}{2 \varOmega \delta^2}, \quad \frac{\kappa_C}{\eta},
		\quad \frac{\kappa_T}{\eta} \to 0,
		\label{cons1}
	\end{equation}
	
	where $\varOmega$ is the angular velocity of rotation. In addition,
	
	\begin{equation}
		\frac{\eta}{2 \varOmega \delta^2} \ll 1
		\label{cons2}
	\end{equation}
	
	has a small but finite value. 
	Depending 
	on whether the thermal and compositional stratifications 
	are stable or unstable, a range of convective 
	regimes is explored in the dynamo simulations
	\cite[and others]{glatzmaier1996anelastic,
		manglik2010dynamo,
		takahashi2014double,
		takahashi2019mercury,
		tassin2021geomagnetic,
		mather2021regimes,
		fan2025dynamos}. 
	Since only unstable density stratification can drive convection, 
	and Earth's outer core is largely unstable in both thermal 
	and compositional gradients, the present study focuses on 
	an unstably stratified core. In this medium,
	a buoyancy
	disturbance under background rotation and a magnetic
	field generates slow and fast Magnetic-Archimedean-Coriolis (MAC)
	waves \cite{brag1967,07bussechapter,jfm21}.
	Dynamo models where the nonlinear inertial force
	is small on the length scale of convection \cite{aditya2022,jfm24}
	indicate that the generation of the slow MAC waves is
	essential for the formation of the axial dipole. 
	
	Strong buoyant forcing in Earth's core is 
	modelled by a finite compositional flux at the
	inner core boundary (ICB) and zero flux
	at the core--mantle boundary (CMB). Low-inertia
	dynamo models with this 
	single-component buoyancy distribution
	indicate that supercritical states of buoyancy
	flux greater than $O(10^2)$ times the critical buoyancy
	flux for onset of convection
	do not support an axial dipole field (see Section \ref{dynamo}
	below). Furthermore,
	the azimuthal polar core flows at the
	dipole--multipole transition are considerably weaker than
	the peak drift of $0.6$--$0.9^\circ$yr$^{-1}$ observed in 
	Earth \cite<see>{hulot2002}. Since much stronger
	compositional buoyancy fluxes would evidently not support
	the axial dipole, compositional
	buoyancy must be paired with relatively weak thermal
	buoyancy to retain the axial dipole character of the
	field as well as the polar drift. In short,
	present-day observational constraints require two-component
	thermochemical convection in the core.
	Although it is well
	known that thermal buoyancy in the core 
	has existed throughout
	Earth's history, how does it stabilize the dipole
	in a two-component convective dynamo? To answer
	this question, the present study begins with
	a strongly driven compositional dynamo which
	inevitably produces a multipolar field and
	introduces thermal buoyancy into it.  
	With the addition of this species, 
	the resultant buoyancy increases and causes
	the generation of additional magnetic field.
	Slow MAC waves are then spontaneously generated,
	which in turn generate a dipole field from
	a chaotic multipolar state. The dipole field
	thus generated is so stable that the compositional
	buoyancy flux must be increased substantially, to
	$O(10^3)$ times the onset buoyancy flux, 
	in order to induce polarity transitions via
	the disappearance of the slow MAC waves. The
	extension of the dipole regime in the two-component
	dynamo relative to the one-component dynamo is 
	fairly predictable
	by linear magnetoconvection satisfying the constraints
	\eqref{cons1} and \eqref{cons2} above. 
	
	The analysis of the
	two-component dynamo is further motivated by the dynamics
	of the mantle.  The present
	study suggests a lower bound for the thermal power ratio,
	the fraction of the total power contributed by thermal
	buoyancy, for the existence of the axial dipole. Above this
	bound, a
	two-component dynamo with homogeneous outer boundary
	heat flux does not
	operate at the threshold of reversals. Rather, it resides deep
	within the dipolar regime. Consequently, a large heterogeneity
	in lower-mantle heat flux would place the dynamo 
	in a polarity-reversing state \cite{arxiv25}.

	This paper is organised as follows. In Section \ref{linear}, 
	the evolution of a localised density perturbation in an 
	unstably stratified fluid is examined in the presence 
	of background rotation, gravity, and a uniform magnetic field 
	in a Cartesian geometry
	corresponding to the ``equatorial radial 
	configuration'' described by \citeA{loper2003buoyancy}.
	Through the analysis of the 
	MAC waves generated by this perturbation, 
	we bring out the role of thermal buoyancy in 
	extending the range of the axial dipole
	field in two-component convection
	relative to one-component convection. 
	In Section \ref{dynamo},  
	%In this section, 
	the role of thermal buoyancy in 
	the dipole--multipole transition is examined by
	the analysis of MAC waves. 
	The generation
	of the dipole in the two-component dynamo is
	shown to follow the same process as that
	of a single component dynamo starting from a small
	seed field, studied earlier by \citeA{aditya2022}.
	Finally, Section \ref{conclusion} 
	summarises the main findings and 
	discusses their implications for 
	the Earth's core.
	
	\section{Evolution of an isolated buoyancy disturbance
		in core-like conditions}
	\label{linear}
	\subsection{Problem set-up and governing equations}
	\label{psetup}
	
	\begin{figure}
		\centering
		\hspace{1.5cm}	\includegraphics[width=0.4\linewidth]{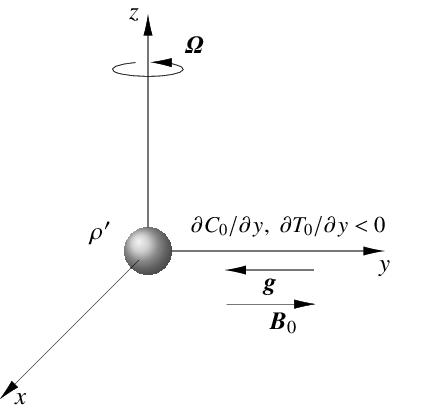}
		\caption{Initial state of a density perturbation 
			$\rho^{\prime}$ in an unstably stratified fluid 
			subject to background rotation 
			$\bm{\varOmega} = \varOmega \hat{\bm{e}}_z$, 
			a uniform magnetic field 
			$\bm{B}_0 = B_0 \hat{\bm{e}}_y$ and gravity 
			$\bm{g} = -g \hat{\bm{e}}_y$, in Cartesian 
			coordinates $(x, y, z)$.}
		\label{coordinate}
	\end{figure}
	
	A localised density disturbance, $\rho^\prime$, arising in an 
	unstably stratified, rotating fluid layer permeated by a 
	uniform magnetic field, is considered. 
	The density perturbation is related to temperature and 
	compositional perturbations, $\Theta$ and $\gamma$, 
	respectively, through
	$\rho^\prime = -\rho (\alpha^T\Theta+\alpha^C\gamma)$, 
	where $\rho$ represents the ambient density, 
	while $\alpha^T$ and $\alpha^C$ denote the 
	coefficients of thermal and compositional expansion, 
	respectively. The initial thermal and compositional 
	perturbations are prescribed in the form
	
	\begin{equation}
		\begin{aligned}
			\{\Theta_{0},\gamma_{0}\} &= 
			A \exp \big[-(x^2+y^2+z^2)/\delta^2 \big],
			\label{pert}
		\end{aligned}
	\end{equation}
	where $A$ represents the perturbation intensity and 
	$\delta$ denotes the characteristic length scale. 
	The initial perturbation, 
	illustrated in Figure \ref{coordinate}, subsequently 
	evolves under the influence of uniform magnetic field 
	$\bm{B}_0 = B_0 \hat{\bm{e}}_y$, gravity 
	$\bm{g} = -g \hat{\bm{e}}_y$, and background 
	rotation $\bm{\varOmega} = \varOmega \hat{\bm{e}}_z$ 
	in Cartesian coordinates $(x,y,z)$.
	
	In an initially quiescent medium, the temperature 
	perturbation given by \eqref{pert} induces a 
	velocity field, $\bm{u}$, which subsequently 
	interacts with the imposed magnetic field, 
	$\bm{B}_0$, generating an 
	induced magnetic field, $\bm{b}$. Both the 
	initial velocity and the initial 
	induced magnetic field are zero. 
	To make the problem dimensionless, 
	lengths are scaled by the perturbation size $\delta$, 
	time is scaled by magnetic diffusion time 
	$\delta^2/\eta$,
	the velocity is scaled by $\eta/\delta$ and
	the magnetic field 
	by $(2\varOmega\rho\mu\eta)^{1/2}$,
	where $\mu$ is the magnetic
	permeability. 
	The temperature is scaled by $\beta^T \delta$ 
	and composition is scaled by $\beta^C \delta$, 
	where $\beta^T$ and $\beta^C$ are the 
	vertical temperature and composition gradients, respectively. 
	Under the Boussinesq approximation, the linearised 
	dimensionless governing equations give the evolution 
	of $\bm{u}$, $\bm{b}$, $\Theta$ and $\gamma$:
	
	\begin{eqnarray}
		&& E_\eta \frac{\partial{\bm{u}}}{\partial{t}} =
		-\nabla{p^\star} +(\nabla\times\bm{b})\times\bm{B}_0
		-\hat{\bm{e}}_z
		\times \bm{u} +Ra_{\ell}^T\Theta \hat{\bm{e}}_y 
		+Ra_{\ell}^C\gamma \hat{\bm{e}}_y +E \nabla^2\bm{u},
		\label{mom}\\
		&& \frac{\partial \bm{b}}{\partial t}  
		= (\bm{B}_0\cdot\nabla)\bm{u}+ \nabla^2 \bm{b},
		\label{ind1} \\
		&& \frac{\partial{\Theta}}{\partial{t}}=
		- \bm{u}_y+ q^T \nabla^2\Theta,
		\label{temp}\\
		&& \frac{\partial{\gamma}}{\partial{t}}=
		- \bm{u}_y+ q^C \nabla^2\gamma,
		\label{comp}\\
		&& \nabla \cdot \bm{u}=\nabla \cdot \bm{b}=0.
		\label{divcond}
	\end{eqnarray}
	
	The dimensionless parameters in equations 
	\eqref{mom}--\eqref{comp}, all based on $\delta$,
	are the Ekman number, $E = \nu / 2\varOmega \delta^2$, 
	the magnetic Ekman number, $E_\eta = \eta / 2\varOmega \delta^2$, 
	and the local compositional and thermal Rayleigh numbers given by 
	$Ra_{\ell}^C = g \alpha^C |\beta^C| \delta^2 / 2 \varOmega \eta$ 
	and $Ra_{\ell}^T = g \alpha^T |\beta^T| \delta^2 / 2 \varOmega \eta$, 
	respectively. In addition, the diffusivity ratios are defined as 
	$q^T = \kappa^T / \eta$ and $q^C = \kappa^C / \eta$. 
	The modified pressure is given by
	$p^\star = p + \frac{1}{2} E_\eta |\bm{u}|^2$.
	
	\subsection{Solutions for the velocity field}
	\label{velsol}
	
	In a rapidly rotating core with
	finite magnetic diffusivity, conditions
	\eqref{cons1} and \eqref{cons2} apply.
	In this limit, a solution of the form
	$\hat{u}_y \propto \mathrm{e}^{\mathrm{i}\lambda t}$ 
	for the coupled equations 
	\eqref{mom}--\eqref{divcond} gives
	the following characteristic equation
	\cite{jfm21,jfm24}:
	
	\begin{equation}
		\begin{aligned}
			&\lambda^5- 2 \mathrm{i}\omega_\eta \lambda^4
			- ({\omega_{A}^C}^2+{\omega_{A}^T}^2+\omega_C^2
			+\omega_\eta^2+2\omega_M^2) \lambda^3\\
			&+2 \mathrm{i} \omega_\eta(
			{\omega_{A}^C}^2+{\omega_{A}^T}^2
			+\omega_C^2+\omega_M^2) \lambda^2\\
			&+\big(({\omega_{A}^C}^2+{\omega_{A}^T}^2)\, \omega_\eta^2
			+\omega_C^2\omega_\eta^2+
			({\omega_{A}^C}^2+{\omega_{A}^T}^2)\, 
			\omega_M^2+\omega_M^4\big) \lambda\\
			&-\mathrm{i} ({\omega_{A}^C}^2+{\omega_{A}^T}^2)\, 
			\omega_\eta\omega_M^2=0,
			\label{char}
		\end{aligned}
	\end{equation}
	the fundamental frequencies 
	in which are given in Table \ref{fundfreq}.
	
	\begingroup
	\setlength{\tabcolsep}{25pt} 
	\renewcommand{\arraystretch}{1.25}
	
	\begin{table}
		\centering
		\caption{Summary of the dimensional and dimensionless 
			forms of the fundamental frequencies. 
			The dimensionless expressions are obtained 
			by scaling the frequencies with $\eta / \delta^2$. 
			Here, $k_h$ denotes the horizontal wavenumber, 
			given by $k_h^2 = k_x^2 + k_z^2$ and 
			$k^2 = k_x^2 + k_y^2 + k_z^2$.	
		}
		\label{fundfreq}
		\begin{tabular}{lcc} 
			\hline
			Frequency & Dimensional & Dimensionless\\
			Inertial wave
			&$\omega_C^2=\dfrac{4\varOmega^2k_z^2}{k^2}$
			&$\dfrac1{E_\eta^2} \dfrac{k_z^2}{k^2}$\\
			Alfv\'en wave
			&$\omega_M^2=\dfrac{(\bm{B}_0 \cdot \bm{k})^2}{\rho\mu}$
			&$\dfrac1{E_\eta} (\bm{B}_0 \cdot \bm{k})^2$\\
			Compositional buoyancy	
			&$-{\omega_{A}^C}^2=g\alpha^C\beta^C\dfrac{k_h^2}{k^2}$
			&$ \dfrac{Ra_\ell^C}{E_\eta} \dfrac{k_h^2}{k^2}$\\
			Thermal buoyancy
			&$-{\omega_{A}^T}^2=g\alpha^T\beta^T\dfrac{k_h^2}{k^2}$
			& $ \dfrac{Ra_\ell^T}{E_\eta}\dfrac{k_h^2}{k^2}$\\
			Magnetic diffusion
			&$\omega_{\eta}^2=\eta^2 k^4$
			&$k^4$\\
			\hline
		\end{tabular}
	\end{table}
	
	\endgroup
	
	For the regime given by the inequality $|\omega_C| \gg |\omega_M|
	\gg |{\omega_{A}^C}|,|{\omega_{A}^T}| \gg |\omega_\eta|$, 
	the approximate roots of
	equation \eqref{char} are \cite{jfm21},

	\begin{eqnarray}
		&&\lambda_{1,2} \approx \pm \bigg(\omega_C
		+ \frac{\omega_M^2}{\omega_C} \bigg) 
		+\mathrm{i} \, \frac{\omega_M^2\omega_\eta}{\omega_C^2}, 
		\label{l12approx}\\ 
		&&\lambda_{3,4} \approx \pm \bigg(\frac{\omega_M^2}{\omega_C}+
		\frac{{\omega_{A}^C}^2+{\omega_{A}^T}^2}{2\omega_C}\bigg) 
		+ \mathrm{i} \, \omega_\eta \,
		\bigg(1-\frac{{\omega_{A}^C}^2+{\omega_{A}^T}^2}{2\omega_M^2}\bigg),
		\label{l34approx} \\
		&&\lambda_{5} \approx \mathrm{i}\, 
		\frac{\omega_\eta({\omega_{A}^C}^2+{\omega_{A}^T}^2)}{\omega_M^2},
		\label{l5approx}
	\end{eqnarray}
	where \eqref{l12approx} and \eqref{l34approx} give
	the damped fast and slow MAC wave frequencies, respectively.
	From \eqref{l34approx}, a resultant buoyancy frequency is defined by
	
	\begin{equation} 
		\omega_A^2={\omega_{A}^C}^2+{\omega_{A}^T}^2, 
		\label{resultant1}
	\end{equation}
	
	and the resultant local Rayleigh number is given by
	
	\begin{equation} 
		Ra_\ell= \frac{g |\alpha^C \beta^C + \alpha^T \beta^T| \delta^2 }
		{2 \varOmega \eta}.
		\label{resultant2}
	\end{equation}
	
	In the limit of 
	$|\omega_C| \gg |\omega_M|$, 
	$|\omega_A|$, the real parts 
	of $\lambda_{3,4}$ are further simplified as

	\begin{equation}\label{eq:slow_root}
		\begin{aligned}
			\mbox{Re}(\lambda_{3,4}) \approx \pm 
			\frac{\omega_M^2}{\omega_C}\,
			\bigg(1+ \frac{{\omega_A}^2}{\omega_M^2} \bigg)^{1/2}.
		\end{aligned}
	\end{equation}

	Therefore, slow MAC waves would be suppressed
	in an unstably stratified fluid where ${\omega_A}^2 < 0$,
	and $|\omega_A| \approx |\omega_M|$.
	
	The solution for $\hat{u}_y$ is given by,

	\begin{equation}
		\hat{u}_y = \sum_{m=1}^{5}D_{m} \,
		\mathrm{e}^{\mathrm{i} \lambda_{m}t},
		\label{solutionyhat}
	\end{equation}

	whereas the solutions for $\hat{u}_x$ 
	and $\hat{u}_z$ are given by,

	\begin{equation}
		\hat{u}_{x} =\hat{u}^H_{x}+\hat{u}^P_{x}
		=\sum_{m=1}^{5}A_{m}
		\mathrm{e}^{\mathrm{i}\lambda_m^H t}
		+ \sum_{m=1}^{5}M_me^{\mathrm{i} \lambda_m^P t},
		\label{solutionxhat}
	\end{equation}

	\begin{equation}
		\begin{aligned}
			\hat{u}_{z} =\hat{u}^H_{z}+\hat{u}^P_{z}
			= \sum_{m=1}^{5}C_{m}
			\mathrm{e}^{\mathrm{i}\lambda_m^H t}
			+ \sum_{m=1}^{5}N_me^{\mathrm{i}\lambda_m^P t},
			\label{solutionzhat}
		\end{aligned}
	\end{equation}

	where the superscripts $H$ and $P$ denote 
	the homogeneous and particular solutions, respectively. 
	The particular solutions are obtained by 
	solving equation \eqref{char} and are 
	approximated in equations 
	\eqref{l12approx}--\eqref{l5approx}. 
	The homogeneous solutions are obtained 
	from equation \eqref{char} 
	by setting $\omega_{A}^C = \omega_{A}^T = 0$.

	The solution for the transforms of the temperature
	and composition perturbations, $\hat{\Theta}$ and  $\hat{\gamma}$,
	are
	
	\begin{equation}
		\hat{\Theta} = \sum_{m=1}^{4}R_{m} 
		\mbox{e}^{\mathrm{i} \lambda_{m}t}, 
		\quad 	\hat{\gamma} = \sum_{m=1}^{4}Q_{m} 
		\mbox{e}^{\mathrm{i} \lambda_{m}t}.
		\label{thetaT}
	\end{equation}

	The coefficients $D_m$, $A_m$, $C_m$, $M_m$, 
	and $N_m$ in equations 
	\eqref{solutionyhat}--\eqref{solutionzhat} 
	are determined from the initial 
	conditions on the velocity and its 
	time derivatives, while the 
	coefficients $R_m$ and $Q_m$ in equation 
	\eqref{thetaT} are obtained from the 
	initial conditions on temperature and 
	composition and their respective time derivatives.
	The procedure followed here is the same
	as in earlier studies \cite{jfm21,jfm24}.
	Since the general solution is a 
	linear superposition of the fast and slow MAC
	wave solutions, the solutions for the two waves
	can be written down individually, as given below:

	\begin{equation}
		\begin{aligned}
			\hat{u}_{x,f}&=M_1 \mathrm{e}^{\mathrm{i}\lambda^P_1 t}+
			M_2 \mathrm{e}^{\mathrm{i}\lambda^P_2 t} + 
			A_{1} \mathrm{e}^{\mathrm{i}{\lambda}_1^H t}+
			A_{2} \mathrm{e}^{\mathrm{i} {\lambda}_2^H t},\\
			\hat{u}_{y,f}& = D_1 \mathrm{e}^{\mathrm{i}\lambda_1 t}
			+ D_2 \mathrm{e}^{\mathrm{i}\lambda_2 t},\\
			\hat{u}_{z,f}&=N_1 \mathrm{e}^{\mathrm{i} \lambda^P_1 t}
			+ N_2 \mathrm{e}^{\mathrm{i} \lambda^P_2 t}
			+C_{1} \mathrm{e}^{ \mathrm{i} {\lambda}_1^H t}
			+C_{2} \mathrm{e}^{\mathrm{i}{\lambda}_2^H t},
			\label{fast}
		\end{aligned}
	\end{equation}
	
	and
	
	\begin{equation}
		\begin{aligned}
			\hat{u}_{x,s}&= M_3 \mathrm{e}^{\mathrm{i} \lambda^P_3 t} 
			+ M_4 \mathrm{e}^{\mathrm{i}\lambda^P_4 t}
			+A_{3} \mathrm{e}^{\mathrm{i}{\lambda}_3^H t} 
			+ A_{4} \mathrm{e}^{\mathrm{i}{\lambda}_4^H t},\\
			\hat{u}_{y,s}& = D_3 \mathrm{e}^{\mathrm{i}\lambda_3 t}
			+D_4 \mathrm{e}^{\mathrm{i}\lambda_4 t},\\
			\hat{u}_{z,s}&=N_3 \mathrm{e}^{\mathrm{i}\lambda^P_3 t}
			+N_4 \mathrm{e}^{\mathrm{i}\lambda^P_4 t}
			+C_{3} \mathrm{e}^{\mathrm{i}{\lambda}_3^H t}+C_{4}
			\mathrm{e}^{\mathrm{i} {\lambda}_4^H t},
			\label{slow}
		\end{aligned}
	\end{equation}
	
	where the subscripts $f$ and $s$ on the left-hand sides of 
	equations \eqref{fast} and \eqref{slow} indicate 
	the fast and slow wave parts of the solution, respectively. 
	The solutions for the induced magnetic field
	can be obtained following a similar approach.
	In two-component convection, the two buoyant species 
	contribute separately to the total power, so  
	the fraction of the total power provided by thermal
	buoyancy is given by
	
	\begin{equation}
		f^T\% = \frac{P^T}{P^T + P^C} \times 100,
		\label{powerrate}
	\end{equation}
	
	where $P^T$ and $P^C$ represent the thermal and 
	compositional power, given by
	
	\begin{equation}
		P^T = \int_V \bm{u}\cdot (Ra_{\ell}^T\Theta 
		\hat{\bm{e}}_y)  \mbox{d}V, \quad
		P^C = \int_V \bm{u}\cdot (Ra_{\ell}^C\gamma 
		\hat{\bm{e}}_y)  \mbox{d}V.
		\label{power}
	\end{equation}
	
	The integrals in \eqref{power} are evaluated within the 
	limits $\pm 20$ in the $(x,y, z)$ space. 
	%This formulation quantifies the extent to which 
	%thermal convection dominates over compositional 
	%convection in thermochemical convection model.
	
	\subsection{MAC waves in thermochemical convection}
	\label{macwtcc}
	
	\begin{figure}
		\centering
		\hspace{-2.5 in}	(a)  \hspace{2.5 in} (b) \\
		\includegraphics[width=0.44\linewidth]{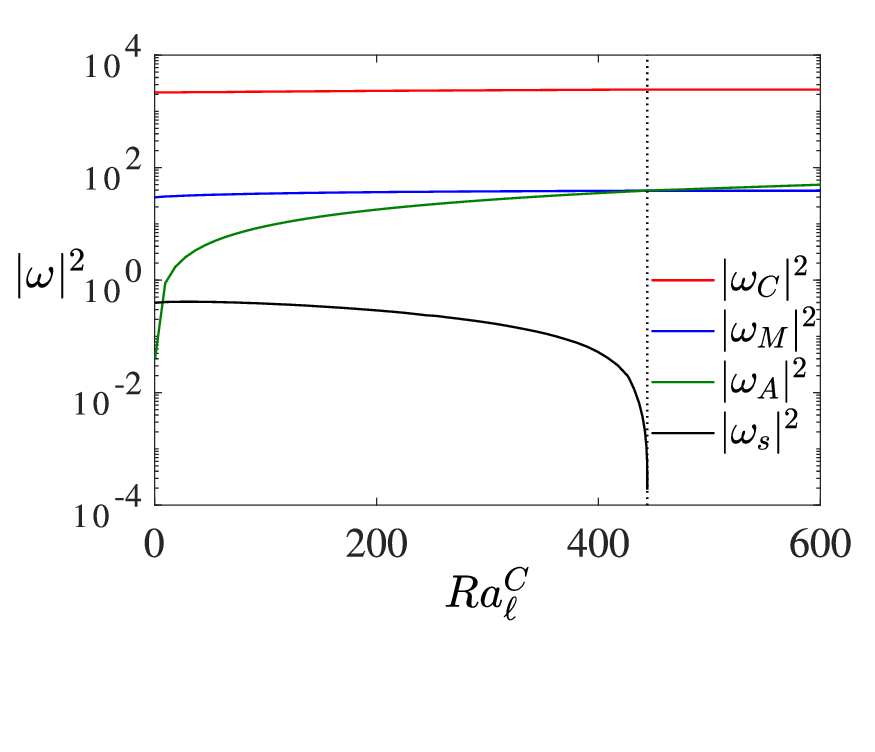}\hspace{0.5cm}
		\raisebox{0.93cm}{\includegraphics[width=0.5\linewidth]{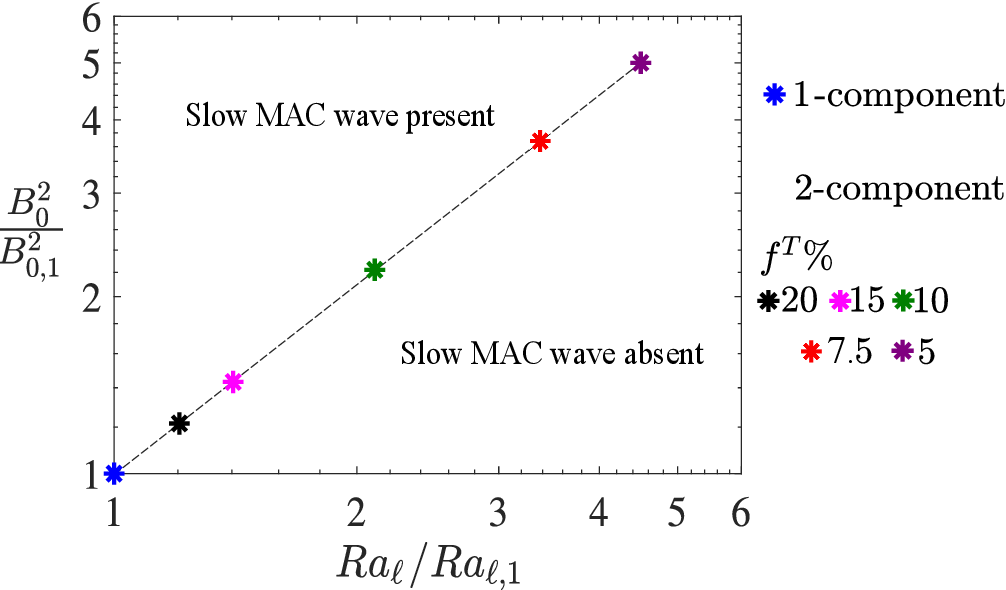}}\\
		\hspace{-2.5 in}	(c)  \hspace{2.5 in} (d) \\
		\hspace*{-1.25cm}\raisebox{0.8cm}{\includegraphics[width=0.48\linewidth]{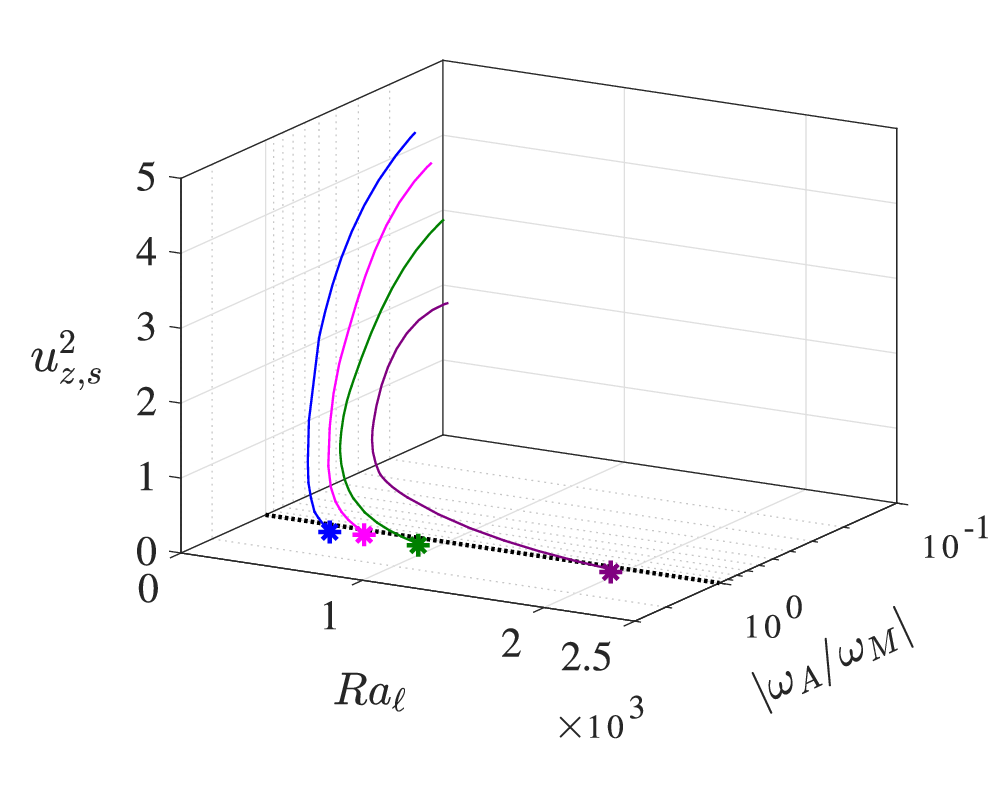}}
		\raisebox{0.8cm}{\includegraphics[width=0.47\linewidth]{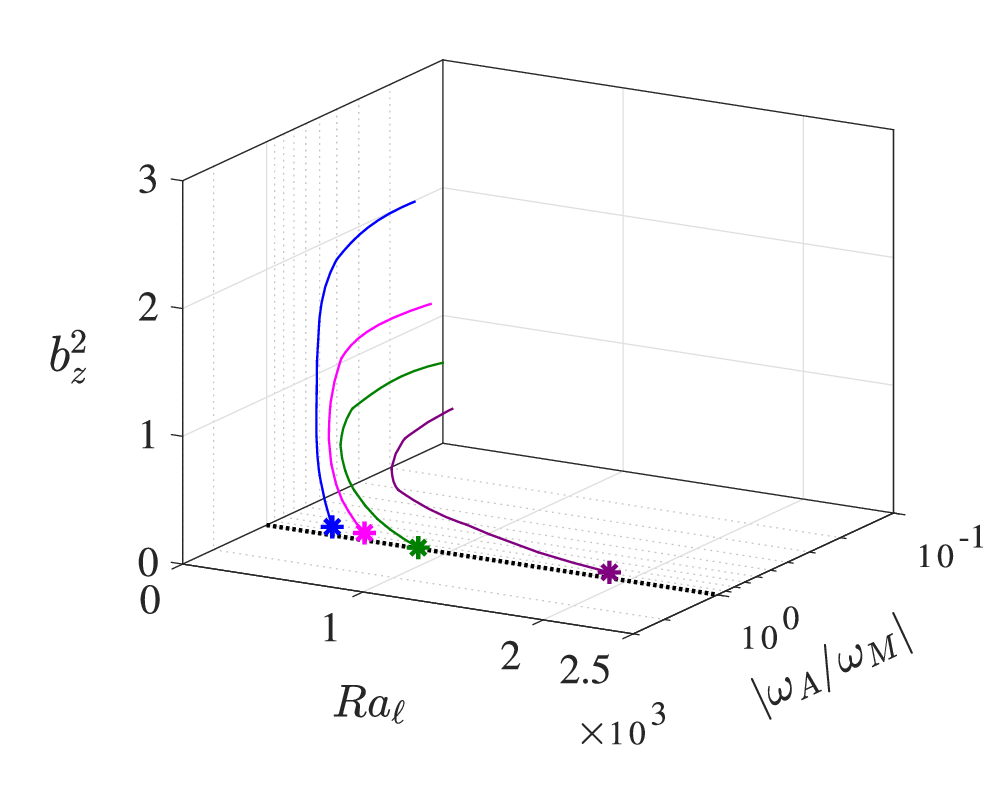}}\\
		\caption{ (a) Absolute values of the squared fundamental 
			frequencies are shown as functions of $Ra_{\ell}^C$ 
			for one-component compositional convection. 
			The dashed vertical line indicates the value of 
			$Ra_{\ell}^C$ at which the slow MAC wave frequency 
			$\omega_s$ approaches zero, corresponding to the 
			condition $\omega_{M} \approx \omega_{A}$. 
			(b) Ratio of $B_0^2$ at the slow MAC wave 
			suppression points for one- and two-component convection, 
			plotted against the ratio of their local Rayleigh numbers. 
			The subscript \lq 1' for $B_0^2$ and $Ra_\ell$ denotes the
			values for one-component compositional convection.
			The thermal power ratios at these suppression points for the 
			two-component models are also indicated.	
			(c) Variation of the square of the axial velocity of the slow 
			MAC waves $u_{z,s}^2$ and 
			(d) square of the axial induced magnetic field $b_{z}^2$ as 
			a function of $Ra_{\ell}$ and $|\omega_{A}/\omega_{M}|$ 
			for one- and two-component convection. The colour codes
			are the same as in (b). 
			The parameters used are $E_\eta = 1 \times 10^{-5}$ 
			and $t/t_\eta = 7.2 \times 10^{-3}$.}
		\label{eklinear}
	\end{figure}
	\begin{figure}
		\centering
		One-component\\
		\hspace*{-0.7cm}(a) $|\omega_A/\omega_M|=0.3$\hspace*{1.5cm}(b) $|\omega_A/\omega_M|=0.96$\\
		\includegraphics[width=0.32\linewidth]{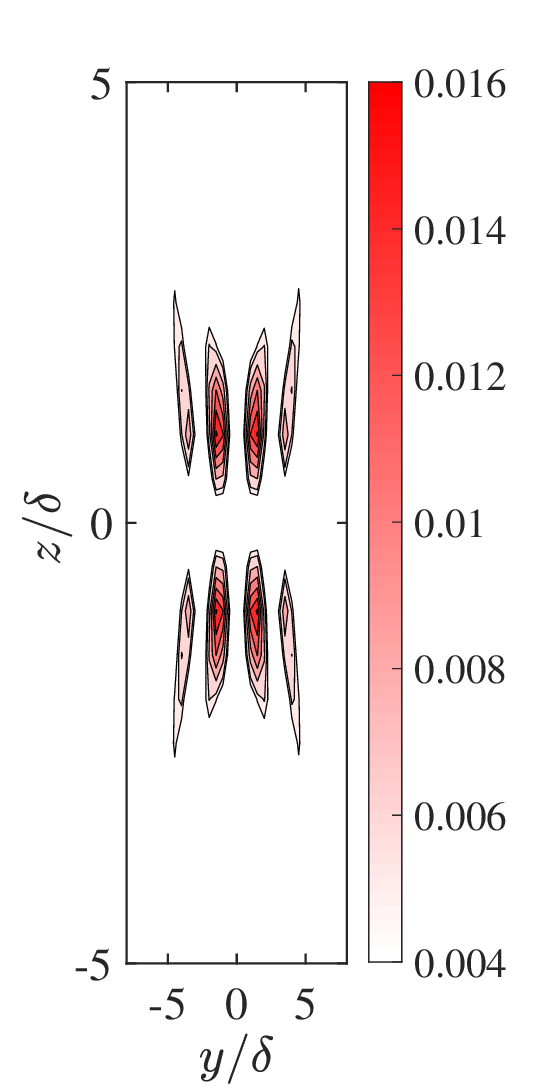}	
		\includegraphics[width=0.32\linewidth]{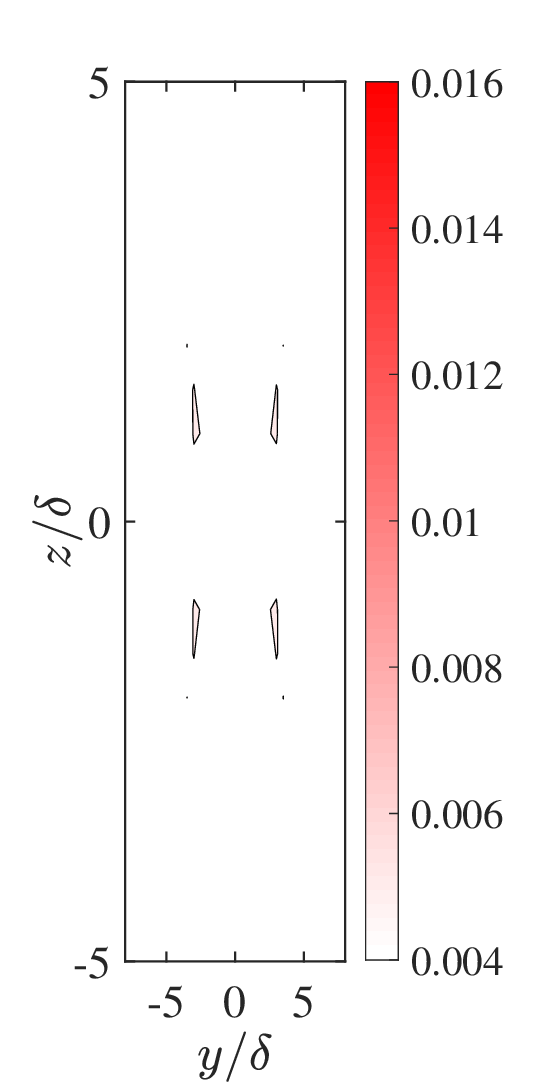}\\
		Two-component\\
		\hspace*{-0.7cm}(c) $|\omega_A/\omega_M|=0.3$
		\hspace*{1.5cm}(d) $|\omega_A/\omega_M|=0.97$\\
		\includegraphics[width=0.32\linewidth]{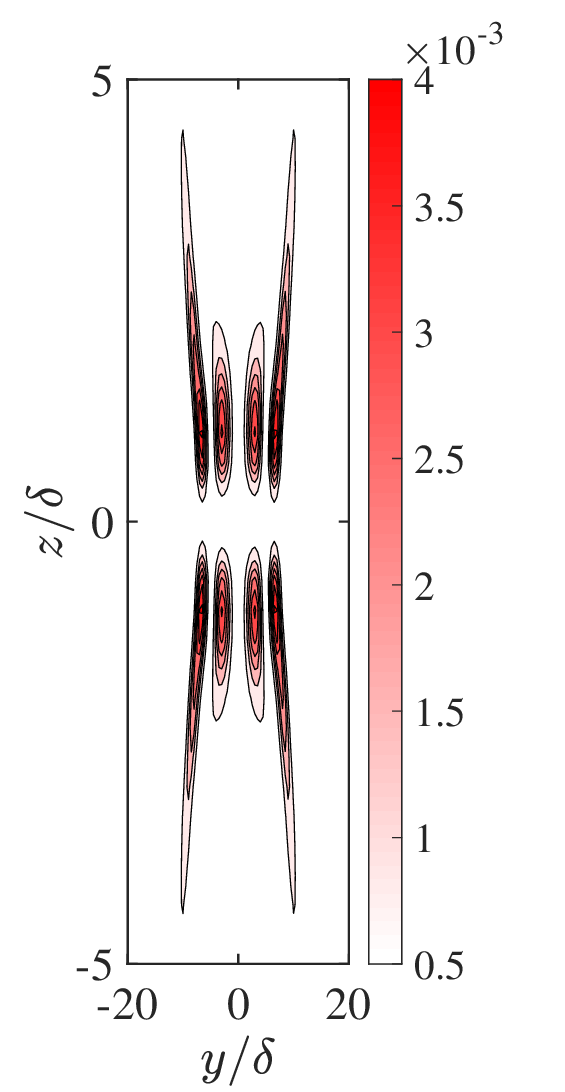}	
		\includegraphics[width=0.32\linewidth]{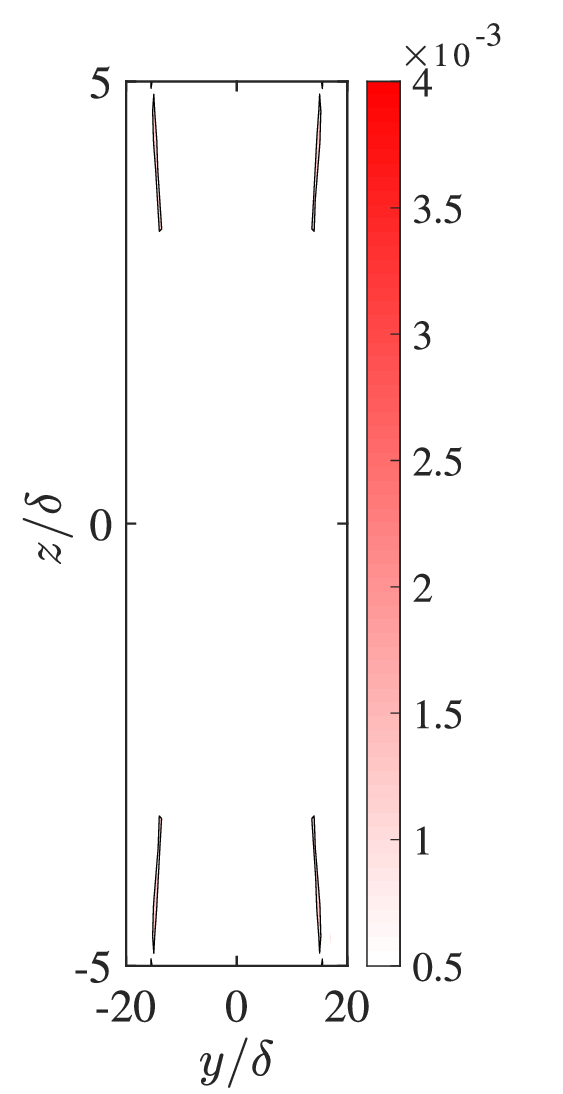}	\\
		\caption{Contour plots of the square of the axial ($z$)
			velocity of the slow ($s$) MAC waves 
			on the $y$--$z$ plane at $x = 0$, 
			for single-component (a--b) and 
			two-component (c--d) convection with 
			$f^T = 5\%$. The corresponding values of 
			$Ra_\ell^C$ ($|\omega_A/\omega_M|$) are 
			(a) 70 (0.3), (b) 440 (0.96), 
			(c) 444 (0.3) and 
			(d) 1900 (0.97). 
			The other parameters are 
			$E_\eta = 1 \times 10^{-5}$ 
			and $t/t_\eta = 7.2 \times 10^{-3}$.}
		\label{uzs}
	\end{figure}

	For the initial value problem described in Section \ref{psetup},
	the parameter space 
	is characterised by the square of the scaled 
	magnetic field, $B_0^2$, 
	and the magnetic Ekman number, $E_\eta$,
	expressible by frequency ratios,
	
	\begin{gather} \refstepcounter{equation} 
		B_0^2 =\dfrac{V_M^2}{2\varOmega\eta}\approx \bigg(\frac{\omega_{M}^2}{\omega_{C}\omega_{\eta}}\bigg)_{\!0}, 
		\quad E_\eta= \dfrac{\eta}{2 \varOmega\delta^2} 
		\approx\bigg(\frac{\omega_{\eta}}{\omega_{C}}\bigg)_{\!0}, 
		\tag{\theequation a,b} \label{ekmanle} 
	\end{gather}
	where $V_M=B_0/\sqrt{\rho\mu}$ is the Alfv\'en wave velocity, 
	and the subscript ‘0’ denotes the initial state of the perturbation. 
	Assuming a plausible ratio of the core depth to the 
	convective length scale, $L/\delta = 10^2$, 
	and choosing a magnetic diffusivity of 
	$\eta = 0.6\,\mathrm{m}^2\,\mathrm{s}^{-1}$ 
	\cite{pozzo2012thermal,jones2015thermal}, 
	the magnetic 
	Ekman number, $E_\eta \approx 1 \times 10^{-5}$.
	In addition, a reasonable estimate for 
	$B_0^2$ is $10^2$, for which the intensity 
	of slow MAC waves becomes comparable to 
	that of the fast MAC waves \cite{jfm21}.
	It may be recalled that this order of magnitude
	corresponds to the square of the peak field intensity
	in self-consistent dipole-dominated
	dynamos \cite{aditya2022}. 
	In addition, the buoyant forcing is measured by 
	the thermal and compositional local Rayleigh 
	numbers, $Ra_\ell^T$ and $Ra_\ell^C$, defined in
	Section \ref{psetup}. One-component 
	(composition-only) convection serves 
	as the starting point for the calculations.
	
	Figure \ref{eklinear}(a) shows the variation of 
	the fundamental frequencies with $Ra_{\ell}^C$. 
	To determine the frequencies, the mean wavenumbers 
	are computed from ratios of $L^2$ norms, for example,
	
	\begin{gather}
		\refstepcounter{equation}
		\bar{k}_x=\frac{||k_x \hat{u}||}{||\hat{u}||},
		\quad \bar{k}=\frac{||k \hat{u}||}{||\hat{u}||},
		\tag{\theequation a,b}
		\label{l2}
	\end{gather}
	which are based on the velocity field. 
	The slow MAC waves exist 
	in the regime characterised by the inequality 
	$|\omega_{C}| > |\omega_{M}| > |\omega_{A}|> |\omega_{\eta}|$. 
	As the forcing is increased, the ratio 
	$|\omega_{A}|/|\omega_{M}|$ tends to unity, 
	leading to the suppression of the slow MAC wave 
	frequency, $\omega_s$. The blue star in figure 
	\ref{eklinear}(b) denotes this suppression point 
	(marked in dashed line in figure \ref{eklinear}(a)) 
	and is shown in a plot of the squared 
	scaled magnetic field ratio versus 
	the local Rayleigh number ratio.
	The values of the squared 
	scaled magnetic field 
	and local Rayleigh number at the
	suppression point for one-component 
	compositional 
	convection are denoted by $B_{0,1}^2$ 
	and $Ra_{\ell,1}$, respectively. 
	An additional thermal buoyancy component 
	is then introduced. To obtain the
	black star in figure \ref{eklinear}(b),
	the thermal power is kept fixed and chosen such that 
	it contributes $\approx$ 20\% of the total convective 
	power when added to the existing compositional state, 
	the value of $Ra_{\ell}^C$ for which 
	suppresses the slow MAC waves 
	in one-component convection. 
	With the addition of this thermal part, the resultant 
	forcing measured by $Ra_\ell$, defined in \eqref{resultant2},
	increases. The increase in the magnitude of the resultant
	buoyancy frequency $\omega_A$, defined in
	\eqref{resultant1}, must be matched by an increased 
	Alfv\'en wave frequency $\omega_M$ 
	for incipient slow MAC wave suppression. 
	This, in turn, implies an increased value of $B_0^2$. 
	For field intensities exceeding this threshold, 
	slow MAC waves exist, whereas for weaker fields they do not exist. 
	The above value of $B_0^2$ thus defines 
	a transitional field strength separating regimes with 
	and without slow MAC waves. 
	
	By keeping $Ra_\ell^T$ fixed while increasing $Ra_\ell^C$, 
	the thermal power ratio $f^T$, defined in \eqref{powerrate}, 
	progressively decreases. Here, the resultant
	local Rayleigh number $Ra_\ell$  
	continues to increase, and at each step a new transitional 
	value of $B_0^2$ can be determined. In this way, the
	colored star points in figure \ref{eklinear}(b)
	are obtained. This numerical experiment demonstrates 
	that the parameter space over which slow MAC waves exist 
	is extended for two-component convection in both $Ra_{\ell}$ 
	and $B_0^2$ relative to one-component convection.

	Figure \ref{eklinear}(c) shows the variation of 
	the square of the axial velocity 
	associated with the slow MAC waves,  $u_{z,s}^2$, with 
	the resultant 
	local Rayleigh number $Ra_\ell$ and with $|\omega_{A}/\omega_{M}|$ for 
	the thermal power ratios investigated
	in figure \ref{eklinear}(b). 
	The axial velocity in real space is 
	obtained from the inverse Fourier 
	transform of $\hat{u}_{z,s}$, 
	
	\begin{equation}
		\mathcal{F}^{-1}\left(\hat{u}_{z,s}\right)=u_{z,s}= 
		{ \frac{1}{(2\pi)^3} } 
		\int_{-\infty}^{\infty}\int_{-\infty}^{\infty}
		\int_{-\infty}^{\infty}\hat{u}_{z,s}
		\mbox{e}^{\mathrm{i} \,\bm{k}\cdot\bm{x}}\, \mathrm{d}k_x\,
		\mathrm{d}k_y\,\mathrm{d}k_z,
		\label{invfourier}
	\end{equation}
	The limits of the wavenumbers 
	in the integral are set to $\pm 3/\delta$, 
	since the values of the initial wavenumbers 
	are $1/\delta$ (see appendix \ref{lmc}). 
	Then, $u_{z,s}^2$ is calculated by squaring the field 
	and summing it over all points in the $(y,z)$ plane at $x=0$.
	It is evident that a larger forcing is required to suppress 
	the slow MAC waves in two-component convection, since both 
	$|\omega_{A}|$ and $|\omega_{M}|$ increase in the presence of 
	the additional buoyancy component. Nevertheless, 
	suppression occurs universally when
	$|\omega_{A}/\omega_{M}| \approx 1$.  
	The suppression of $u_{z,s}^2$ in 
	one- and two-component convection 
	is illustrated in figure~\ref{uzs}.
	Using a similar procedure, the 
	square of the induced axial 
	magnetic field $b^2_z$ is plotted
	in figure \ref{eklinear}(d). Interestingly,
	$b_z^2$ also vanishes as the slow 
	MAC waves are suppressed, which indicates that
	the slow MAC waves have a direct bearing 
	on axial field generation while the 
	fast MAC waves have no role in the generation
	of the axial field (see also figure S1 in the Supporting Information).

	The linear magnetoconvection model 
	provides the motivation for 
	investigating the role of thermal buoyancy in 
	extending the dipole--multipole transition in the nonlinear 
	dynamos, where the 
	disappearance of the slow MAC waves is shown to 
	correlate with magnetic polarity transitions, as in
	one-component dynamos \cite{jfm24}.
	In Section \ref{dynamo} below, we shall see that the peak magnetic field 
	intensity in the dynamo increases by the
	addition of thermal buoyancy of a fixed power, 
	whereby the suppression of the slow MAC waves
	occurs at a much higher compositional Rayleigh number
	relative to that in one-component purely compositional
	convection. However, since there would be an upper bound
	for the intensity of the 
	field generated in a rotating two-component dynamo, the
	large increase of $|\omega_A|$ at high compositional
	Rayleigh numbers would not be accompanied by a commensurate
	increase in $|\omega_M|$, because of which 
	one anticipates a lower bound
	of the thermal power ratio for the existence of an axial
	dipole.
	\section{Nonlinear dynamo simulations}
	\label{dynamo}
	
	A thermochemically driven dynamo is considered within an 
	electrically conducting fluid confined between two 
	concentric, co-rotating spherical surfaces corresponding to 
	the inner core boundary (ICB) and the core-mantle boundary (CMB). 
	The ratio of the inner radius $r_i$ to the outer radius 
	$r_o$ is set to 0.35. Lengths are scaled by $L=r_o-r_i$ 
	and the time 
	is scaled by $L^2/\eta$, where $\eta$ is the magnetic diffusivity. 
	The velocity is scaled by $\eta/L$ and the magnetic 
	field is scaled by $(2\varOmega \mu\eta\rho)^{1/2}$, 
	where $\varOmega$ is the angular velocity of rotation, 
	$\mu$ is the magnetic permeability, and $\rho$ is the fluid density. 
	The temperature is scaled by $\beta^T L$ and the composition is 
	scaled by $\beta^C L$, where $\beta^T$ and $\beta^C$ are
	the mean thermal and compositional 
	gradients within the shell, respectively. 
	Under the Boussinesq approximation, the governing 
	non-dimensional magnetohydrodynamic (MHD) equations for 
	the velocity, 
	magnetic field, temperature, and composition are as follows:
	
	\begin{align}
		E Pm^{-1}  \Bigl(\frac{\partial {\bm u}}{\partial t} + 
		(\nabla \times {\bm u}) \times {\bm u}
		\Bigr)+  {\hat{\bm{z}}} \times {\bm u} = - \nabla p^\star  
		+ Ra^T \, T \, {\bm r} \,  \nonumber\\  
		+ Ra^C \, C \, {\bm r} +(\nabla \times {\bm B})
		\times {\bm B} +E\nabla^2 {\bm u}, \label{momentumdd} \\
		\frac{\partial {\bm B}}{\partial t} = 
		\nabla \times ({\bm u} \times {\bm B}) 
		+ \nabla^2 {\bm B},  \label{inductiondd}\\
		\frac{\partial T}{\partial t} +({\bm u} \cdot \nabla) T =  
		Pm Pr^{-1} \,
		\nabla^2 T+S_o,  \label{heat1dd}\\
		\frac{\partial C}{\partial t} +({\bm u} \cdot \nabla) C =  
		Pm Sc^{-1} \,
		\nabla^2 C+S_i,  \label{comp1dd}\\
		\nabla \cdot {\bm u}  =  \nabla \cdot {\bm B} = 0.
		\label{divdd}
	\end{align}
	
	The modified pressure, $p^*$, in equation \eqref{momentumdd}, 
	is expressed as $p + \frac{1}{2} E \, Pm^{-1} \, |\bm{u}|^2$. 
	The dimensionless parameters governing the system are 
	the Ekman number $E = \nu/2\varOmega L^2$ the Prandtl number 
	$Pr = \nu/\kappa^T$, the Schmidt number $Sc = \nu/\kappa^C$; 
	the magnetic Prandtl number, $Pm = \nu/\eta$ and 
	the modified thermal and compositional Rayleigh numbers
	$Ra^T = g \alpha^T \beta^T L^2/2\Omega\eta$ and 
	$Ra^C = g \alpha^C \beta^C L^2/2\Omega\eta$, respectively. 
	Here, $g$ is the gravitational acceleration, $\nu$ 
	is the kinematic viscosity, $\kappa^T$ and $\kappa^C$ 
	are the thermal and compositional diffusivities 
	and $\alpha^T$ and $\alpha^C$ are the coefficients of thermal 
	and compositional expansion, respectively.
	
	The velocity field satisfies the no-slip condition at the boundaries 
	while the magnetic field is subject to electrically insulating conditions. 
	The basic-state superadiabatic 
	temperature profile represents internal 
	heating with a source $S_0$, representing secular cooling. 
	An isothermal boundary condition is imposed at the ICB, whereas a 
	fixed heat flux condition is applied at the CMB. The 
	basic-state temperature profile is given by
	
	\begin{equation}
		T_0 = \frac{Pr}{Pm} \frac{S_0}{6} (r_i^2 - r^2).
	\end{equation}
	
	For composition, a uniform volumetric sink $S_i$ is considered, 
	with a constant flux at the ICB and zero flux at the CMB. 
	The basic-state compositional gradient is given by
	\begin{equation} 
		\frac{\partial C_0}{\partial r} = 
		\frac{Sc}{Pm} \frac{S_i}{3} \bigg(\frac{r_0^3}{r^2} - r\bigg).
	\end{equation}
	The numerical calculations are performed using a pseudospectral 
	code that employs spherical harmonic expansions in the 
	angular coordinates $(\theta, \phi)$ and finite-difference 
	discretisation in the radial direction \cite{07willis}.
	
	In this study, two Ekman numbers are considered, 
	with $Pm = Sc$ chosen to ensure that the local Rossby number 
	$Ro_\ell$ is always $<0.1$. This condition ensures 
	that the dynamo simulations remain in a regime dominated by rotation, 
	with small nonlinear inertial effects. The mean spherical harmonic 
	degrees associated with convection and energy injection, 
	denoted by $l_C$ and $l_E$ respectively, are 
	given by
	
	\begin{equation}
		l_C = \dfrac{\sum l \hspace{1pt} E_k(l)}{\sum E_k(l)}, \hspace{10pt}
		l_E = \dfrac{\sum l \hspace{1pt} E_T(l)}{\sum E_T(l)}. \label{elldef} 
	\end{equation}
	
	Here, $E_k(l)$ gives the kinetic energy spectrum
	while $E_T(l)$ gives the spectrum obtained from 
	the product of the transform of $u_r T$ and its conjugate. 
	The total kinetic and magnetic energies in the saturated 
	dynamo state are determined by volume integration,
	
	\begin{equation} 
		E_k = \dfrac{1}{2} \int \bm{u}^2 \mbox{d}V, \quad E_m 
		= \dfrac{Pm}{2E} \int \bm{B}^2 \mbox{d}V. 
		\label{energies}
	\end{equation}
	
	The relative dipole field strength, $f_{dip}$, is given by 
	the ratio of the mean dipole field strength to the total field 
	strength in harmonic degrees $l = 1$--$12$ at the outer boundary, 
	following \citeA{chraub2006}. In all dipole-dominated cases, 
	this value is greater than 0.35.
	The fraction of the total power provided by
	thermal buoyancy in the two-component system is 
	given by
	
	\begin{equation}
		f^T = \frac{P^T}{P^T + P^C},
		\label{ft} 
	\end{equation}
	
	where
	
	\begin{equation}
		P^T = Ra^T \int_{V} T \bm{u} \cdot \bm{r}  \mbox{d}V, \quad P^C 
		= Ra^C  \int_{V} C \bm{u} \cdot \bm{r}  \mbox{d}V.
		\label{ftc}
	\end{equation}
	The key time-averaged outputs from the dynamo simulations 
	are summarised in Table \ref{tablepara}. The resolutions
	used in the simulations 
	and the computed wavenumbers are given in Table S1,
	Supporting Information.
	
	\begingroup
	\setlength{\tabcolsep}{1.75pt}
	\renewcommand{\arraystretch}{1}
	
	\captionsetup{
		width=\linewidth,
		justification=justified,
		singlelinecheck=false
	}
	
	\begin{longtable}{*{16}{c}}
		\caption{Summary of the main input and output 
			parameters of the one- and two-component 
			dynamo simulations considered in this study. Here,
			$Ra^C$ and $Ra^T$ denote the modified compositional 
			and thermal Rayleigh numbers, respectively, and
			$Ra_{\mathrm{cr}}^{C}$ and 
			$Ra_{\mathrm{cr}}^{T}$ are the respective 
			critical Rayleigh numbers for the onset of 
			non-magnetic convection. The thermal power ratio $f^T$ 
			is defined in equation \eqref{ft}, 
			$Rm$ is the magnetic Reynolds number,
			$Ro_\ell$ is the local Rossby number,
			$E_k$ and $E_m$ are the 
			time-averaged total kinetic and magnetic energies
			per unit mass 
			respectively, $B^2_{peak}$ is the square of 
			the peak magnetic field in the saturated dynamo, 
			$B^2_{rms}$ is the root mean square
			value of the square 
			magnetic field measured in the spherical shell,
			$u_{\phi,sc}$ is the scaled peak magnitude of the 
			time and azimuthally averaged $\phi$ velocity inside the
			tangent cylinder (in $^\circ \mbox{yr}^{-1}$), and $f_{dip}$
			is the relative dipole field strength. 
			In addition, $\omega_C$, $\omega_M$, 
			and $\omega_A$ are the linear inertial wave frequency, 
			Alfv\'en wave frequency and buoyancy frequency, 
			respectively, and are defined in Table \ref{fundfreq}. 
			Types \lq D', \lq R', and \lq M' denote dipolar, reversing, 
			and multipolar dynamos, respectively.}
		\label{tablepara}\\
		
		\toprule
		$Ra^C$&$\dfrac{Ra^C}{Ra_{\mathrm{cr}}^{C}}$&$Ra^T$
		&$\dfrac{Ra^T}{Ra_{\mathrm{cr}}^{T}}$
		&$f^{T}$&$Rm$
		&$Ro_\ell$&$\dfrac{E_m}{E_k}$&$B^2_{peak}$
		& $B^2_{rms}$&$|\omega_{M}^2|$
		&$\bigg|\dfrac{\omega_{A}}{\omega_{M}}\bigg|$
		&$\bigg|\dfrac{\omega_{M}}{\omega_{C}}\bigg|$
		&$u_{\phi,sc}$
		&$f_{dip}$&Type\\
		&&&&&&&&&&$\times10^9$&&&&&D/R/M\\
		\midrule
		\endfirsthead
		
		\toprule
		$Ra^C$&$\dfrac{Ra^C}{Ra_{\mathrm{cr}}^{C}}$&$Ra^T$
		&$\dfrac{Ra^T}{Ra_{\mathrm{cr}}^{T}}$
		&$f^{T}$&$Rm$
		&$Ro_\ell$&$\dfrac{E_m}{E_k}$&$B^2_{peak}$
		& $B^2_{rms}$&$|\omega_{M}^2|$
		&$\bigg|\dfrac{\omega_{A}}{\omega_{M}}\bigg|$
		&$\bigg|\dfrac{\omega_{M}}{\omega_{C}}\bigg|$
		&$u_{\phi,sc}$
		&$f_{dip}$&Type\\
		&&&&&&&&&&$\times10^9$&&&&&D/R/M\\
		\midrule
		\endhead
		
		\bottomrule
		\multicolumn{16}{r}{{Continued on next page}} \\
		\endfoot
		
		\bottomrule
		\multicolumn{16}{r}{{End of the table}} \\
		%	\label{tablepara}
		\endlastfoot
		\multicolumn{16}{c}{$E = 6 \times 10^{-5},Pm=5,Sc=5,Pr=0.5$}\\
		
		\multicolumn{16}{c}{ Pure composition}\\
		1000 &50&0&0&0&112&0.005&4.57&32&0.68&0.11&0.80&0.24&0.12&0.91&D\\
		2000 &100&0&0&0&152&0.008&4.40&45&1.05&0.23&0.75&0.37&0.25&0.86&D\\
		2500 &125&0&0&0&173&0.009&4.21&47&1.07&0.25&0.80&0.40&0.34&0.84&D\\
		2750 &137&0&0&0&177&0.010&4.06&48&1.12&0.44&0.81&0.19&0.45&0.82&D\\
		3000 &150&0&0&0&234&0.012&0.42&20&0.22&0.13&1.27&0.28&0.13&0.21&R\\
		4000 &200&0&0&0&267&0.014&0.33&08&0.17&0.04&2.69&0.16&0.17&0.17&M\\

		\multicolumn{16}{c}{ Pure thermal}\\
		
		%	0&0&220&1&100&\multicolumn{11}{c}{Onset (nonmagnetic) thermal Rayleigh number}\\
		0&0&440&2&100&151&0.005&12.35&71&2.98&0.46&0.25&0.54&0.05&0.75&D\\
		
		\multicolumn{16}{c}{Two-component convection}\\
		
		1000 &50&220&1&36.8&133&0.007&11.27&40 &1.94&0.26&0.56&0.41&0.11&0.88&D\\
		3000 &150&220&1&21.3&200&0.010&6.99 &71 &3.05&0.46&0.68&0.56&0.29&0.79&D\\	
		6000 &300&220&1&17.3&281&0.015&3.77 &110&3.14&0.73&0.75&0.68&0.31&0.75&D\\
		8000 &400&220&1&15.1&317&0.017&2.78 &130&3.22&0.97&0.76&0.86&0.60&0.72&D\\	
		9000 &450&220&1&14.0&327&0.018&2.65 &145&3.47&1.08&0.77&0.91&0.63&0.69&D\\	
		12000&600&220&1&12.4&373&0.021&2.16 &172&3.56&1.33&0.80&0.96&0.65&0.65&D\\	
		14000&700&220&1&10.2&426&0.023&1.90 &126&3.62&1.23&0.84&0.87&0.71&0.64&D\\
		15000&750&220&1&8.4&443&0.025&1.15&113&2.87&1.11&0.94&0.78&0.75&0.61&D\\
		16000&800&220&1&7.4 &461&0.027&0.92 &105&1.53&0.98&1.06&0.83&0.30&0.27&R\\
		17000&850&220&1&6.8 &475&0.028&0.83 &100&1.11&0.75&1.26&0.74&0.28&0.20&R\\
		18000&900&220&1&6.2 &483&0.029&0.71 &95 &1.08&0.73&1.31&0.72&0.27&0.17&M\\
		20000&1000&220&1&4.1 &550&0.032&0.31 &80 &1.06&0.72&1.47&0.61&0.24&0.14&M\\
		\multicolumn{16}{c}{ }\\
		%	3000&150&0&0&0&234&0.012&0.42&20&0.22&0.13&1.27&0.28&0.13&0.21&R\\
		%	3000&150&50 &0.23&5.5&235&0.012& 0.50& 23& 0.25&0.16&1.08&0.29&0.15&0.30&R\\
		%	3000&150&75 &0.34&7.8&240&0.012&0.55&26&0.27&0.19&1.05&0.32&0.18&0.34&R\\
		%	3000&150&100&0.45&9.5&220&0.011& 1.24& 42& 0.76&0.25&0.85&0.38&0.19&0.62&D\\
		%	3000&150&150&0.68&14.8&205&0.010& 5.20& 53& 2.48&0.36&0.71&0.42&0.25&0.83&D\\
		3000&150&220&1.00&21.3&200&0.010&6.99 &71 &3.05&0.46&0.68&0.56&0.29&0.79&D\\	
		3000&150&440&2.00&45.2&231&0.011& 7.89&102& 4.97&0.55&0.62&0.52&0.35&0.78&D\\
		3000&150&500&2.27&48.7&234&0.011&10.26&131& 6.79&0.74&0.55&0.61&0.37&0.72&D\\
		3000&150&660&3.00&57.1&274&0.013&10.32&171& 8.45&0.97&0.48&0.69&0.43&0.68&D\\
		3000&150&750&3.41&64.2&290&0.013&10.43&182& 9.85&1.09&0.46&0.73&0.48&0.62&D\\
		3000&150&1000&4.54&74.4&340&0.014&9.79&196&12.08&1.11&0.48&0.74&0.50&0.60&D\\
		\multicolumn{16}{c}{ }\\
		9000&450&440&2&24.2&356&0.018&3.02 &181&4.12&1.16&0.73&0.95&0.86&0.73&D\\	
		\multicolumn{16}{c}{ }\\
		\multicolumn{16}{c}{ $E = 1.2 \times 10^{-5},Pm=2,Sc=2, Pr=0.2$}\\
		\multicolumn{16}{c}{ Pure composition}\\
		1000 &35.7&0&0&0&144&0.005&3.64&14&0.43&0.24&0.76&0.21&0.14&0.94&D\\
		3000 &107.1&0&0&0&255&0.012&3.49&33&1.01&0.63&0.77&0.42&0.25&0.85&D\\
		6000 &214.3&0&0&0&316&0.012&3.11&58&1.72&1.01&0.85&0.53&0.41&0.80&D\\
		9000 &321.4&0&0&0&410&0.017&2.44&69&1.81&1.49&0.88&0.69&0.48&0.72&D\\
		10500&375.0&0&0&0&454&0.017&2.34&73&1.91&1.36&0.95&0.62&0.55&0.65&D\\
		12000&428.6&0&0&0&534&0.025&0.53&27&0.61&0.98&1.33&0.67&0.23&0.13&R\\
		
		\multicolumn{16}{c}{Two-component convection}\\
		12000&428.6&500&2.4&22.6&422&0.015&8.16&181&8.65&2.34&0.51&0.052&0.70&0.80&D\\ 
		16000&571.4&500&2.4&21.3&530&0.018&6.38&228&8.79&3.52&0.55&0.058&0.75&0.72&D\\ 
		20000&714.3&500&2.4&18.5&547&0.022&5.41&273&8.84&3.99&0.60&0.064&0.79&0.79&D\\ 
	\end{longtable}
	%\end{landscape}
	\endgroup
	\subsection{The effect of thermal buoyancy on
		a two-component convective dynamo}
	\label{thermalrole}
	\begin{figure}
		\centering
		\hspace{-2.5 in}	(a)  \hspace{2.5 in} (b) \\
		\includegraphics[height=0.35\linewidth]{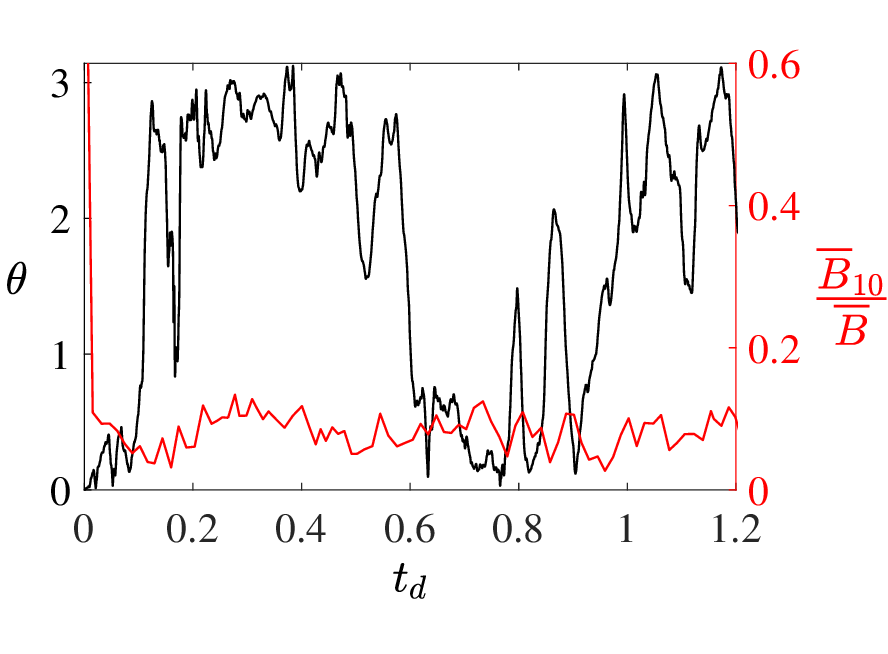}
		\includegraphics[height=0.35\linewidth]{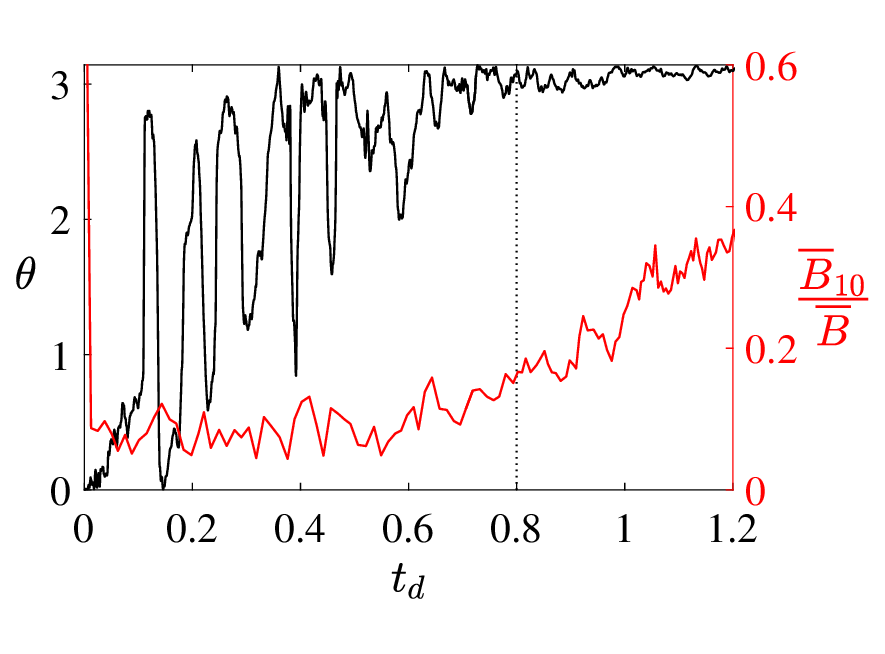}	\\
		\hspace{-5.35 in}	(c)  \\
		\includegraphics[height=0.32\linewidth]{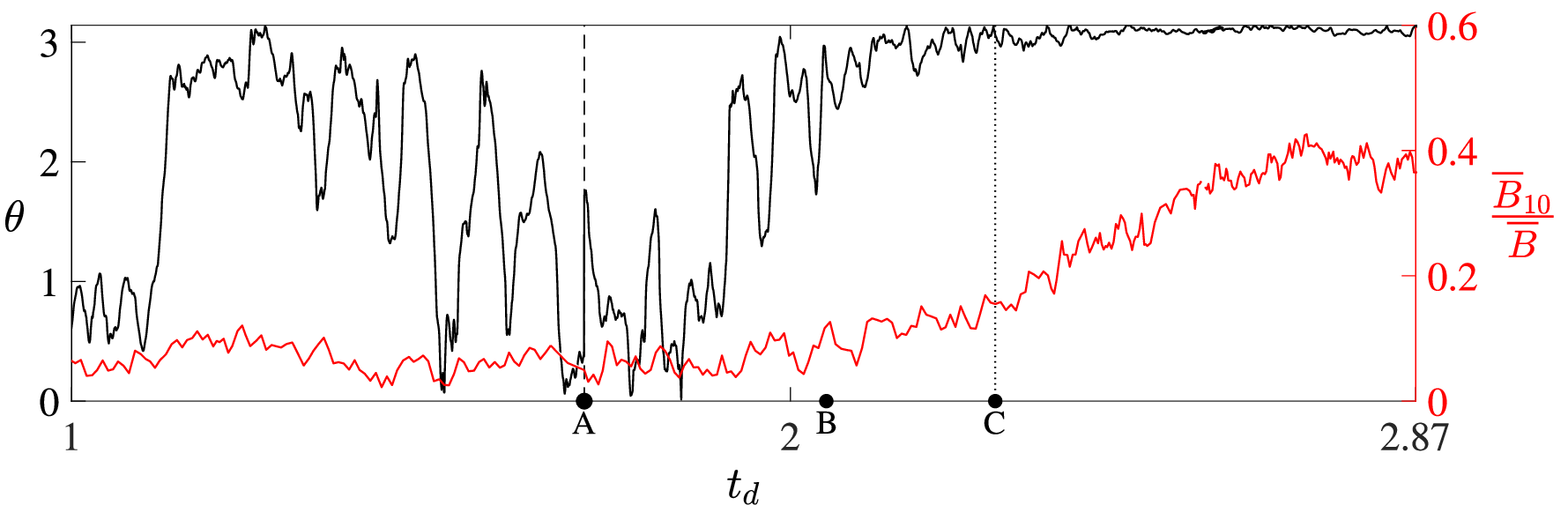}	\\
		\hspace{-1.2 in}	(d) \hspace{0.75 in} Time `B' \hspace{1.15 in} 
		(e) \hspace{0.65 in}  Time `C' \\
		\includegraphics[height=0.2\linewidth]{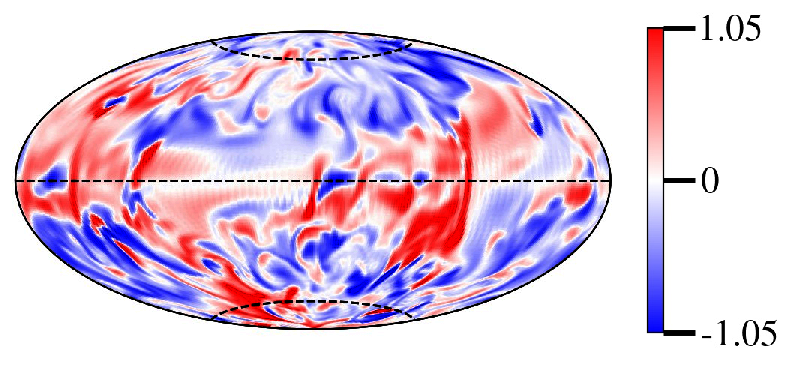}
		\includegraphics[height=0.2\linewidth]{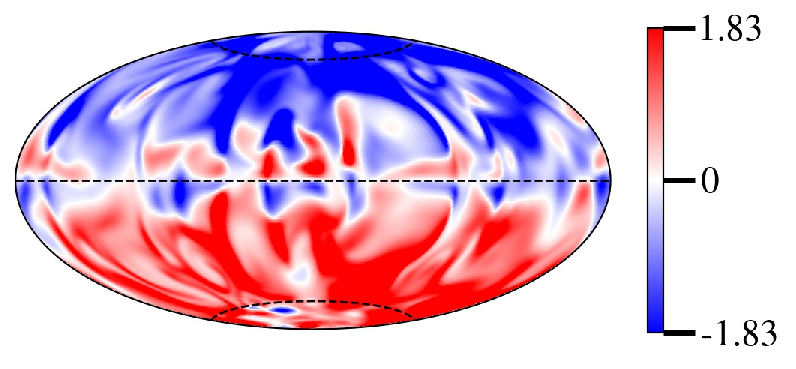}\\
		\caption{Evolution of the dipole colatitude 
			and the ratio of the root 
			mean square value of the axial magnetic field 
			($\bar{B}_{10}$) to 
			the root mean square value of the total 
			magnetic field ($\bar{B}$) 
			with magnetic diffusion time. 
			Panels (a) and (b) show two dynamo simulations initiated 
			from a seed magnetic 
			field for (a) $Ra^T = 0$, $Ra^C = 3000$ and 
			(b) $Ra^T = 220$, $Ra^C = 3000$. 
			Panel (c) corresponds to a simulation 
			initiated from a reversing state with 
			$Ra^T = 0$, $Ra^C = 3000$, where at time 
			\lq A' (dashed vertical line) 
			thermal buoyancy with $Ra^T = 220$ is introduced, 
			contributing $\approx 20\%$ of the 
			total convective power. 
			The dotted vertical lines in 
			panels (b) and (c) indicate the time of 
			dipole formation. (d) \& (e): Contours of 
			the radial magnetic field at the outer 
			boundary at time \lq B' and \lq C' 
			are shown. The dynamo 
			parameters are $E = 6 \times 10^{-5}$, 
			$Pm = Sc = 5$, and $Pr = 0.5$.}
		\label{tiltfdip}
	\end{figure}
	
	\begin{figure}
		\centering
		\hspace{-1.5 in}(a)  \hspace{1.5 in} (b)  \hspace{1.5 in} (c)  \\
		\includegraphics[width=0.3\linewidth]{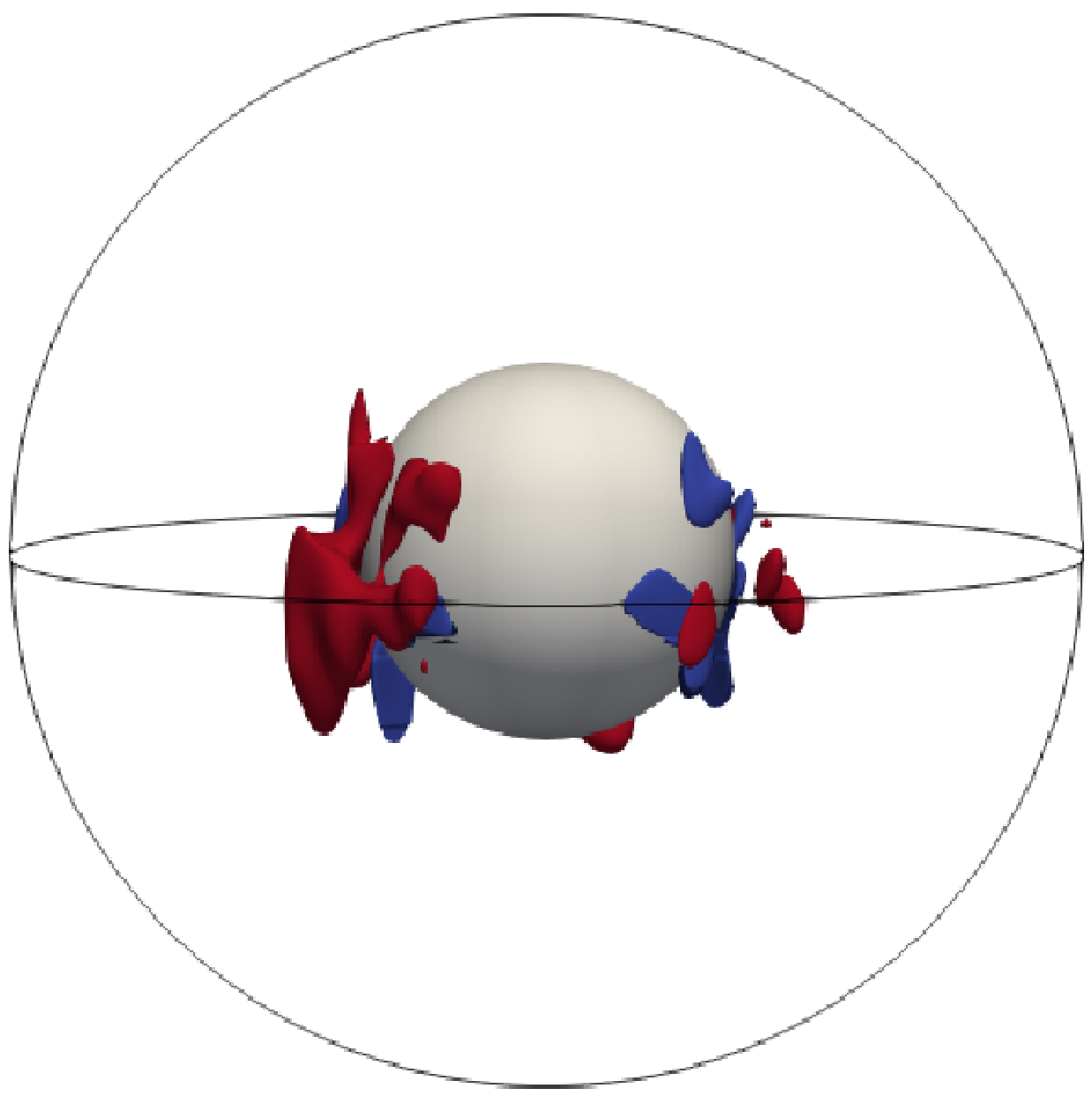}	\hspace*{0.1cm}
		\includegraphics[width=0.3\linewidth]{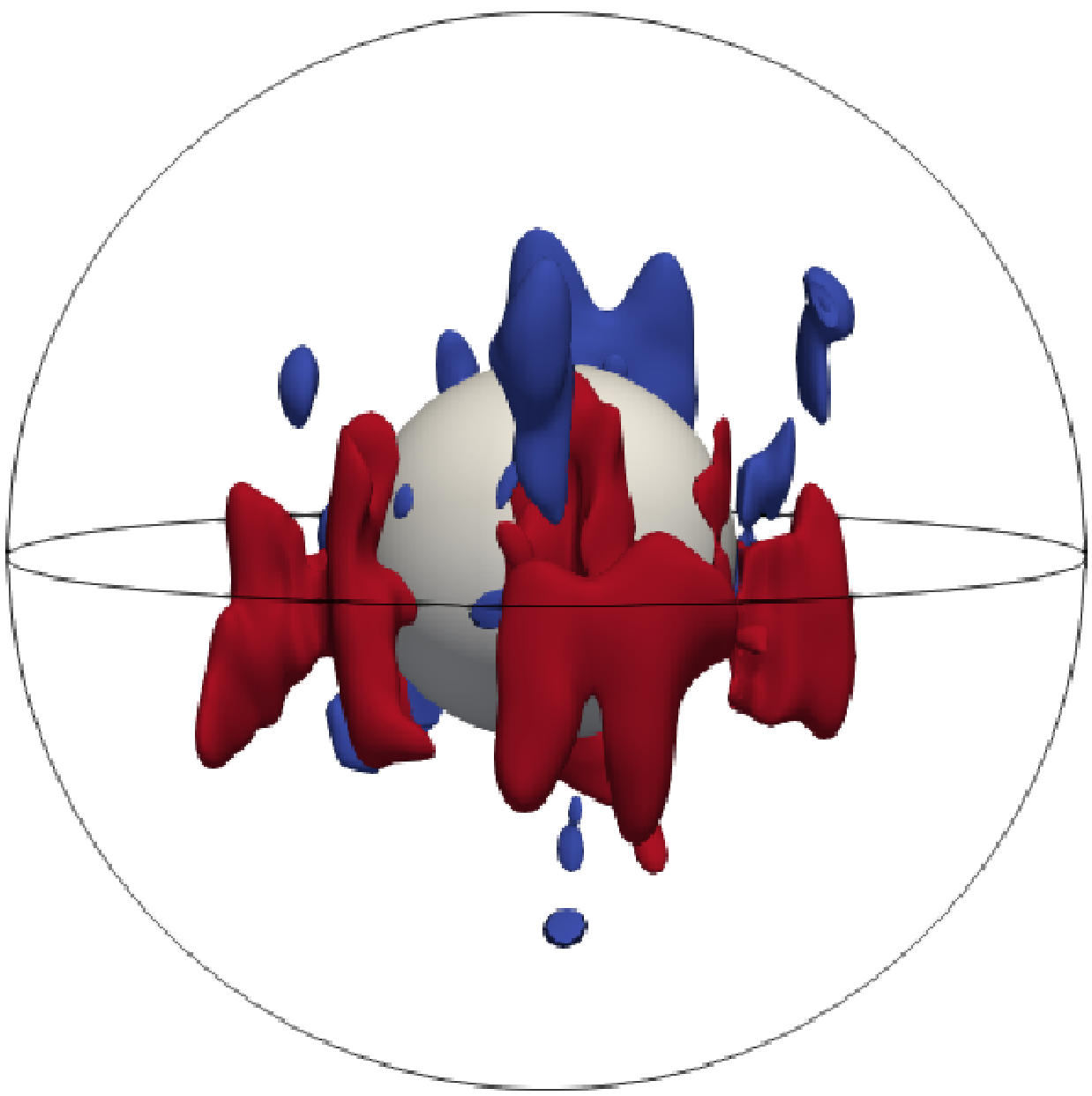}	\hspace*{0.1cm}
		\includegraphics[width=0.3\linewidth]{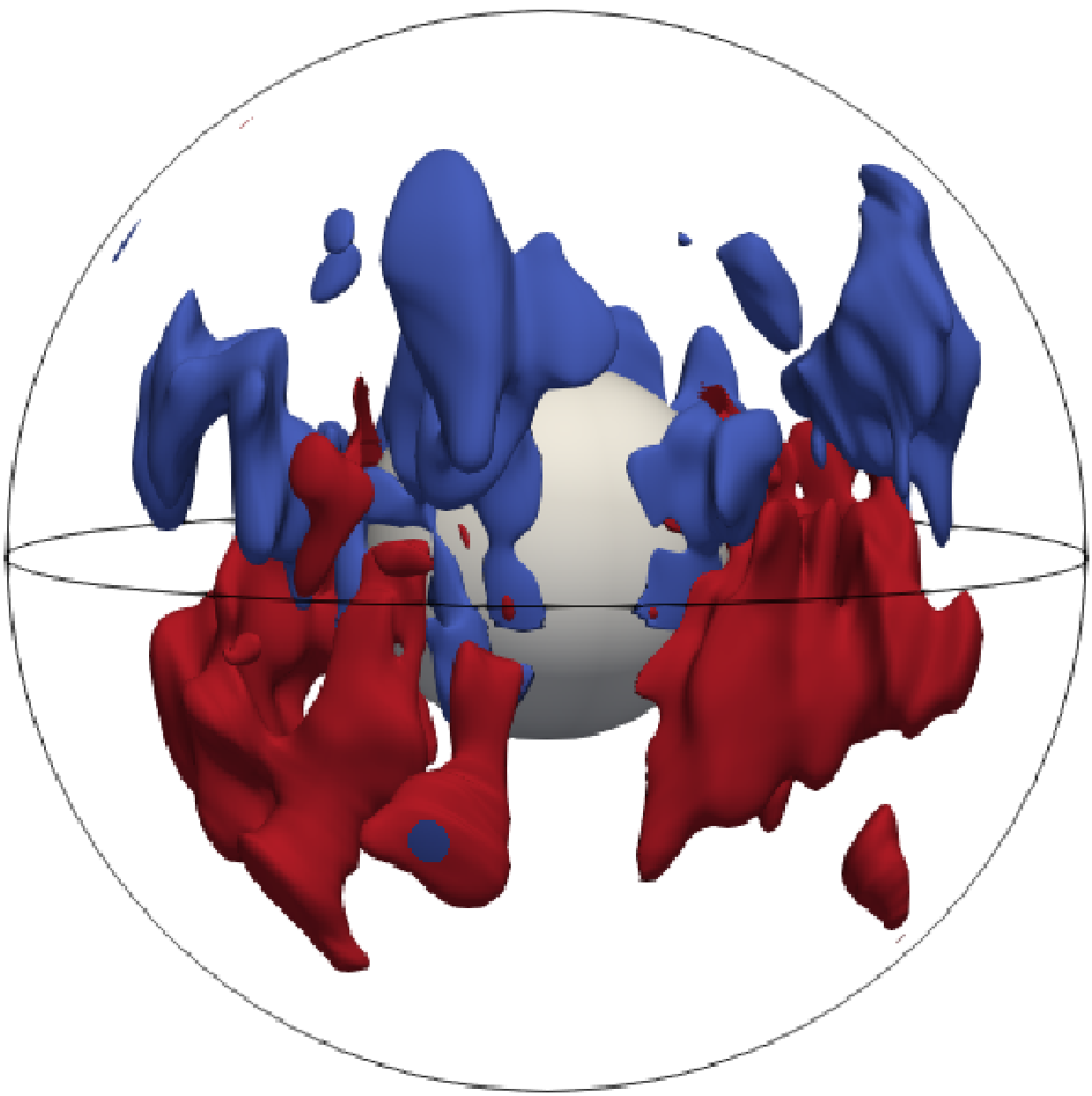}
		\caption{Helicity isosurfaces (contour level 
			$\pm 7 \times 10^5$) at times \lq A', \lq B' and \lq C' 
			(marked in figure \ref{tiltfdip} c)
			for $l \leq 17$. The other dynamo parameters are 
			$Ra^T=220$, $Ra^C=3000$, 
			$E = 6 \times 10^{-5}$ and $Pm = Sc = 5$.}
		\label{helicity}
	\end{figure}
	%%%
	This study examines the role of thermal buoyancy in the 
	dipole--multipole transition in a 
	two-component convective dynamo.
	We begin the study with one-component simulations driven 
	solely by a compositional gradient. As the
	Rayleigh number $Ra^C$ is progressively increased, 
	the dipole-dominated dynamo transitions to
	a polarity-reversing state at $Ra^C = 3000$ and to
	a multipolar state at higher $Ra^C$.  
	In the dipolar runs, the peak
	azimuthal speeds of the polar circulation scaled up to Earth's core, 
	denoted by $u_{\phi, sc}$ and calculated as in
	\citeA{grl2005}, are considerably
	weaker than the observed
	peak drift of 0.6--0.9 $^\circ \mbox{yr}^{-1}$
	\cite{hulot2002} (Table \ref{tablepara}). Therefore,
	compositional buoyancy is combined with thermal
	buoyancy to realize the axial dipole field and
	the magnetically controlled polar vortices within the
	tangent cylinder. 
	%Two distinct cases are considered. 
	%In the first case, dynamo simulations are initiated 
	%from a seed magnetic field driven by compositional buoyancy, 
	%with and without the of thermal buoyancy. 
	%In the second case, a compositional convective state 
	%exhibiting polarity reversals is considered. 
	%A thermal Rayleigh number contributing 
	%approximately 20\% of the convective power is then 
	%imposed on this reversing state, and the subsequent evolution 
	%of the thermochemical convection is analysed.
	
	Figure~\ref{tiltfdip} presents the evolution of the 
	colatitude of the axial dipole field, $\theta$, at the outer 
	boundary (left axis), together with the ratio of the root mean 
	square value of the axial magnetic field $\bar{B}_{10}$ 
	to the root mean square value of the total magnetic field 
	$\bar{B}$ (right axis). The colatitude of the axial dipole
	field is determined from the Gauss coefficients of the 
	spherical harmonic expansion, as given by
	\begin{equation}
		\cos\theta = g_1^0/|\bm{m}|, \quad 
		\bm{m} = (g_1^0, g_1^1, h_1^1),
		\label{diplat}
	\end{equation}
	
	where $g_1^0$, $g_1^1$, and $h_1^1$ are obtained 
	from the Schmidt-normalised expansion of the scalar 
	potential of the magnetic field \cite[pp. 142--143]{glatz2013}.
	%Figure~\ref{tiltfdip}(a--b) illustrates the 
	%initialisation of the simulation 
	%with a small seed magnetic field. 
	In figure~\ref{tiltfdip} (a), the simulation is driven 
	solely by a compositional gradient at $Ra^C = 3000$, 
	for which the dynamo exhibits 
	field reversals. 
	In figure \ref{tiltfdip} (b), thermal buoyancy
	with $Ra^T = 220$, 
	which contributes $\approx$ 20\% of the total convective power, 
	is introduced alongside the same compositional 
	buoyancy as in (a). Here, $\bar{B}_{10}$ is 
	found to increase and 
	the dipole is eventually stabilised, as 
	indicated by the colatitude.
	In Figure~\ref{tiltfdip}(c), the simulation is initiated from a 
	reversing state driven by compositional 
	buoyancy with $Ra^C = 3000$. 
	At a stage marked  \lq A' in 
	figure~\ref{tiltfdip}(c), thermal 
	buoyancy contributing 20\% of the total 
	convective power is introduced. 
	Following this, $\bar{B}_{10}$ intensifies 
	and saturates. 
	Point \lq C' marks the time of dipole formation. 
	The radial magnetic fields at the 
	outer boundary at times \lq B' and \lq C' in
	figures \ref{tiltfdip} (d) and (e) show the field 
	configurations before and after the formation 
	of the axial dipole.
	
	The generation of kinetic helicity in
	energy-containing scales $l \le l_E$ of the dynamo is a useful
	measure of axial dipole formation \cite{prf18,aditya2022}. 
	The helicity isosurface 
	plots at the energy-containing scales at 
	times \lq A', \lq B', and \lq C' (marked in 
	figure~\ref{tiltfdip}(c)) are shown in 
	figures~\ref{helicity}(a--c).
	
	A dynamo driven by purely by thermal buoyancy
	of intensity $Ra^T= 440$ ($2 \times$ its value for
	convective onset) produces
	a strong-field dipolar dynamo with $B_{rms}^2 \approx 3$ and
	$B_{peak}^2 \approx 70$, although its polar circulation
	scaled up to the core,
	$u_{\phi,sc} =0.05 ^\circ \mbox{yr}^{-1}$, is weak. Adding
	compositional buoyancy of intensity $Ra^C= 9000$ ($450 
	\times$ its value for onset) gives a dipolar
	dynamo with $B_{rms}^2 \approx 4$ and
	$B_{peak}^2 \approx 180$, and
	$u_{\phi,sc} =0.86 ^\circ \mbox{yr}^{-1}$ (Table \ref{tablepara}),
	consistent with
	observation. While a two-component dynamo is not
	essential for dipole formation itself, 
	that fact that it places
	the dynamo deep within the dipolar regime helps explain
	the role of substantial lower-mantle heat flux variations in inducing
	occasional polarity reversals in Earth (Section \ref{conclusion}).

	\subsection{The role of slow MAC waves in 
		axial dipole formation in two-component dynamos}
	\label{slowmac}
	
	\begin{figure}
		\centering
		\hspace{-5 in}(a)   \\
		\hspace*{-0.1cm}\includegraphics[width=1\linewidth]
		{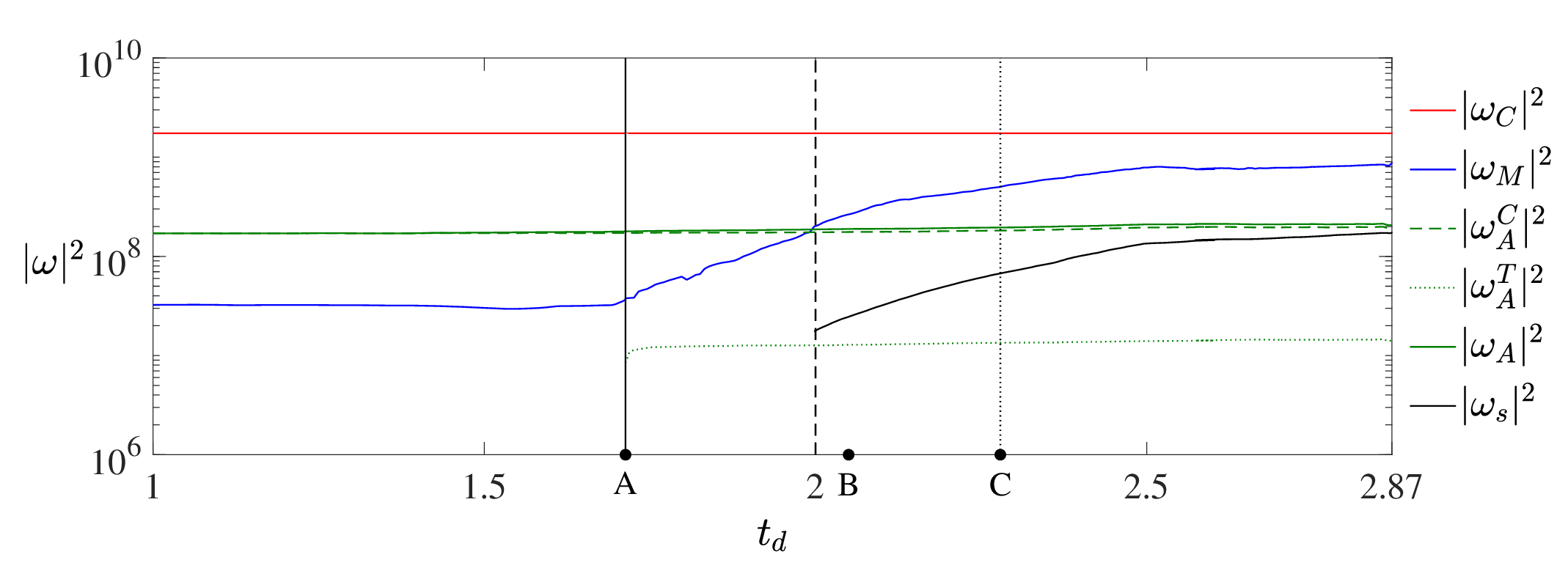}\\
		{\large Time \lq A'}\\
		\hspace{-2.35 in}	(b)  \hspace{2.35 in} (c) \\
		\includegraphics[width=0.45\linewidth]{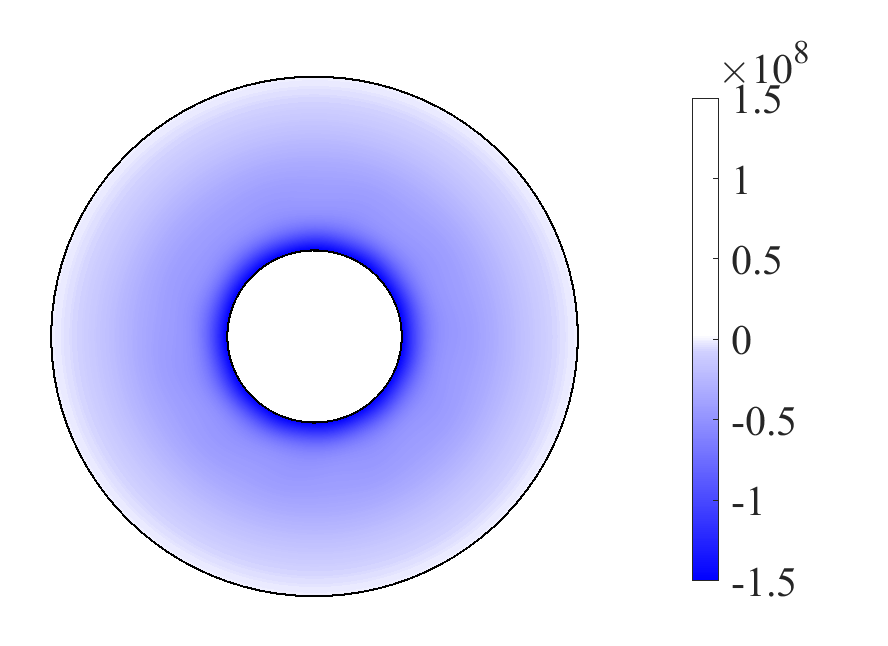}
		\includegraphics[width=0.45\linewidth]{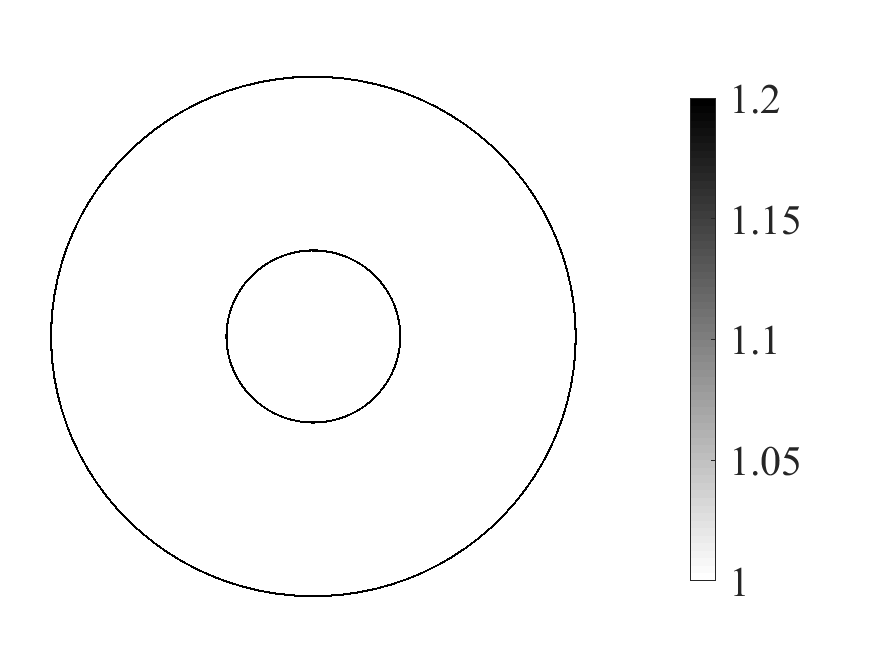}\\
		{\large Time \lq C'}\\
		\hspace{-2.35 in}	(d)  \hspace{2.35 in} (e) \\
		\includegraphics[width=0.45\linewidth]{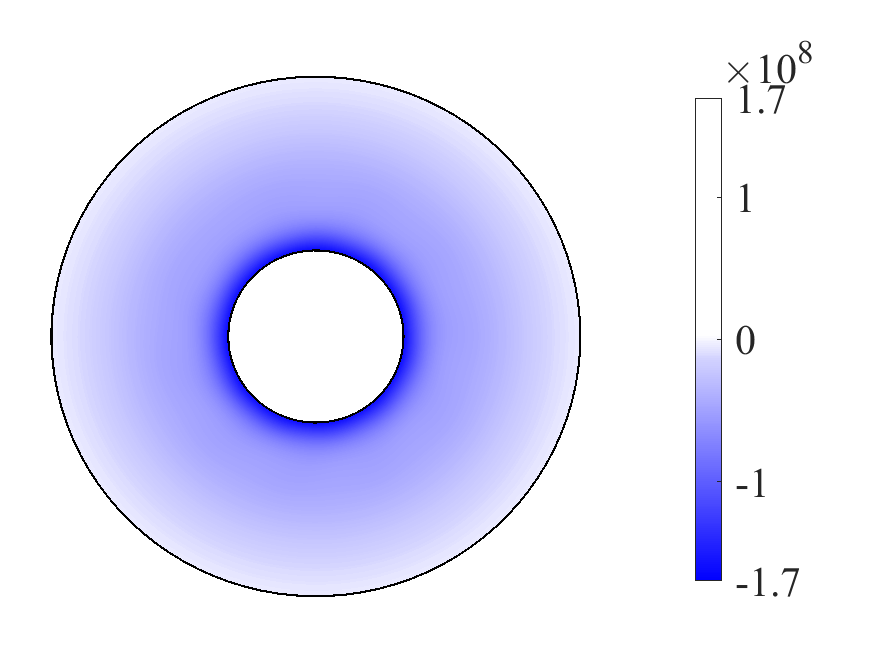}
		\includegraphics[width=0.45\linewidth]{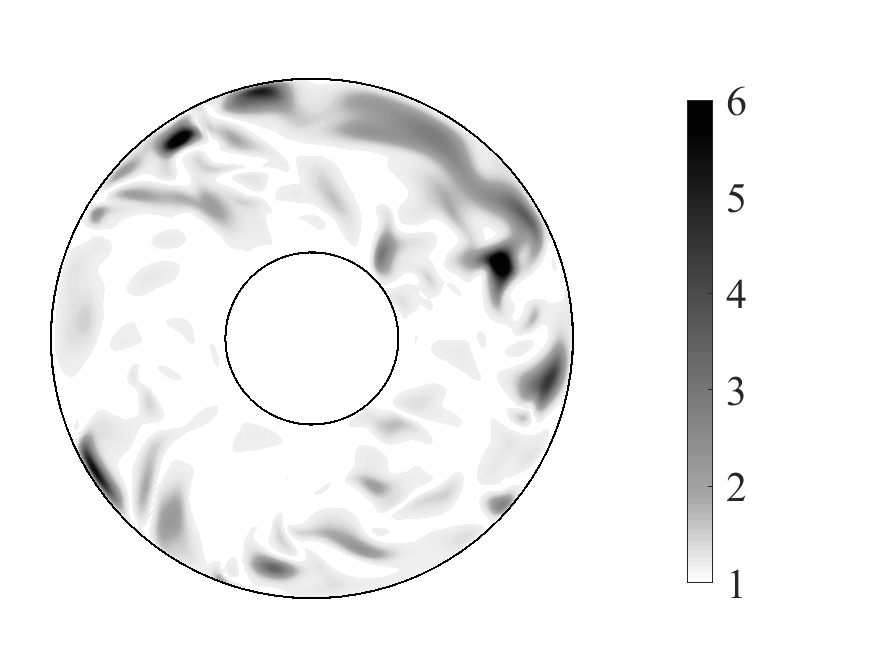}\\
		\caption{(a) The evolution of fundamental frequencies 
			in the dynamo over time is shown. The simulation begins 
			with purely compositional convection 
			($Ra^T = 0$, $Ra^C = 3000$) which 
			belongs to the reversing regime.
			At point \lq A' (dashed line), thermal
			buoyancy close to its onset ($Ra^T = 220$) 
			is introduced into
			the simulation. At point \lq C' (dotted line), 
			the stable axial dipole forms. Point 
			\lq B' is an intermediate state where 
			slow MAC waves are present, but the field
			is non-dipolar.
			Horizontal section 
			plots at height $ z=0.2 $ below the equator showing  
			$ \omega_{A}^2 $ (b, d) 
			and $|\omega_{M}^2/\omega_{A}^2|$ (c, e)
			for (b, c) $Ra^T=0, Ra^C=3000$ and 
			(d, e) $Ra^T=220, Ra^C=3000$. 
			Figures (b) \& (c) correspond to
			point \lq A' whereas (d) \& (e)
			correspond to point \lq C' in figure \ref{tiltfdip}(c).
			The other dynamo parameters 
			are  $E = 6 \times 10^{-5} $, 
			$Pm = Sc = 5,Pr = 0.5$.}	
		\label{freqtseries}
	\end{figure}
	\begin{figure}
		\centering
		\hspace{-1.75 in}	(a)  \hspace{1.5 in} (b)  \hspace{1.5 in} (c) \\
		\includegraphics[width=0.32\linewidth]{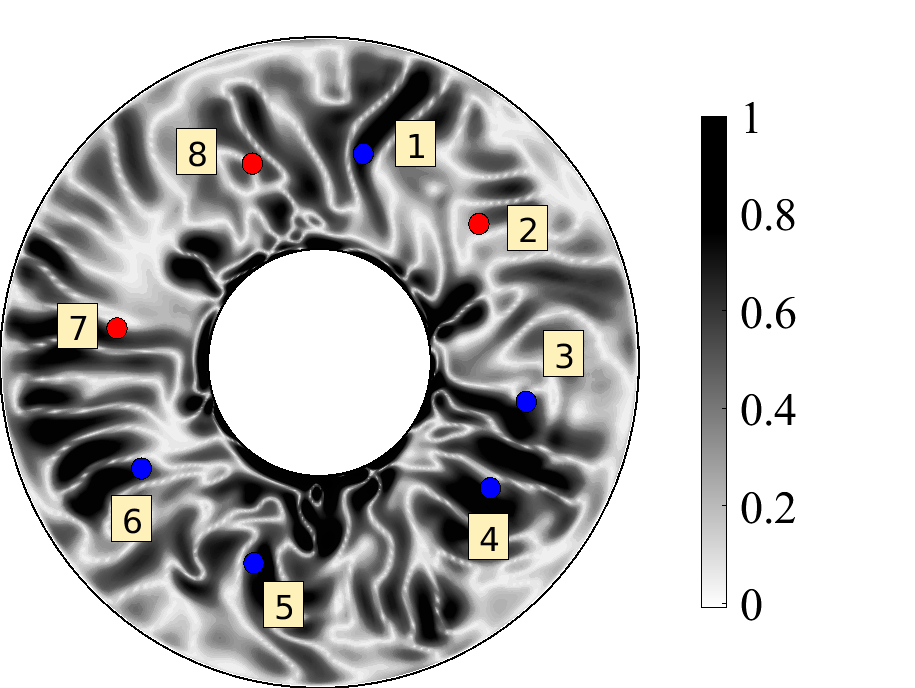}
		\includegraphics[width=0.32\linewidth]{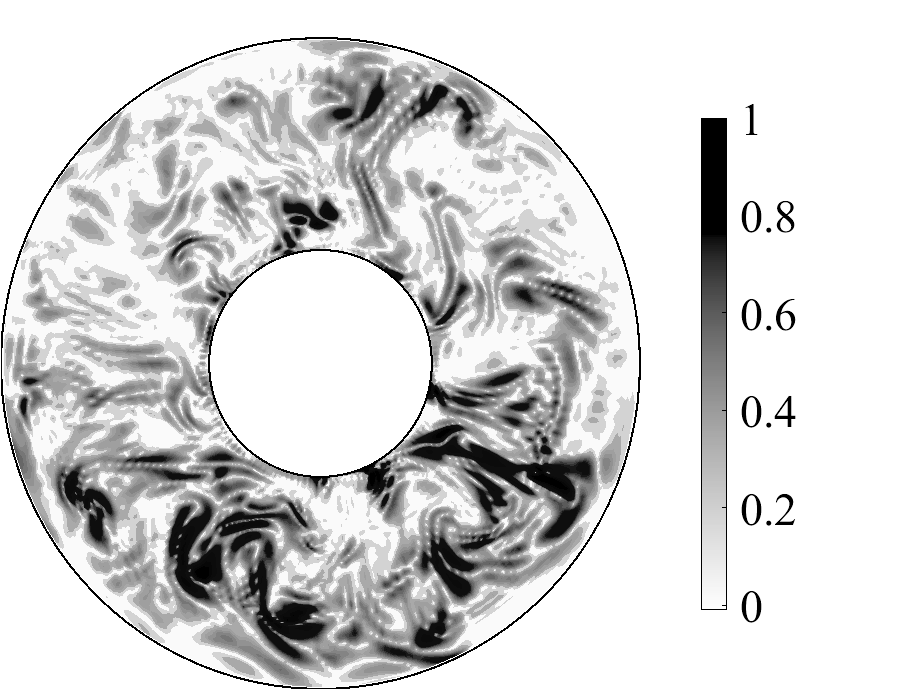}
		\includegraphics[width=0.32\linewidth]{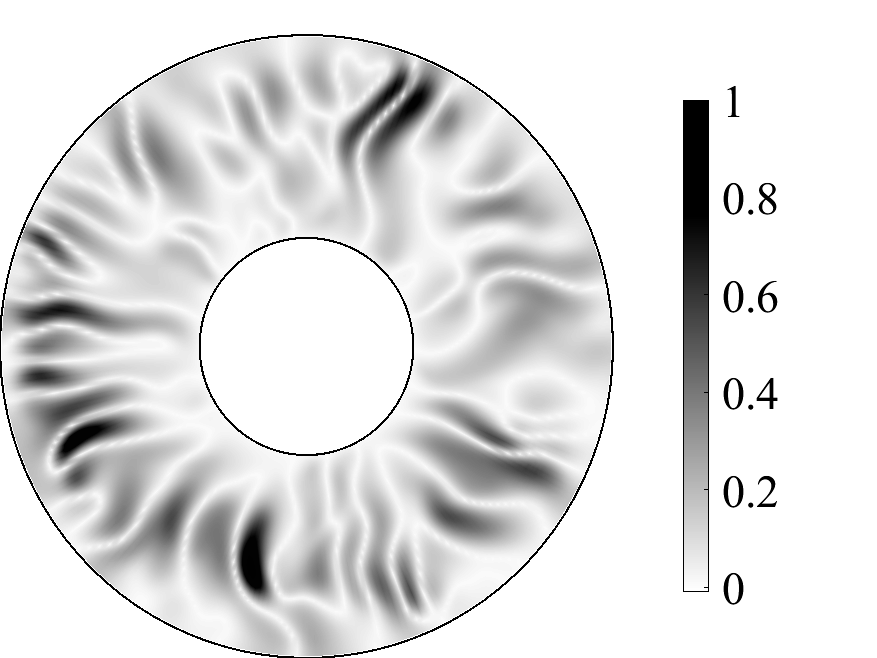}	\\
		\hspace{-1.75 in}	(d)  \hspace{1.5 in} (e)  \hspace{1.5 in} (f) \\
		\hspace*{-0.75cm}
		\includegraphics[width=0.32\linewidth]{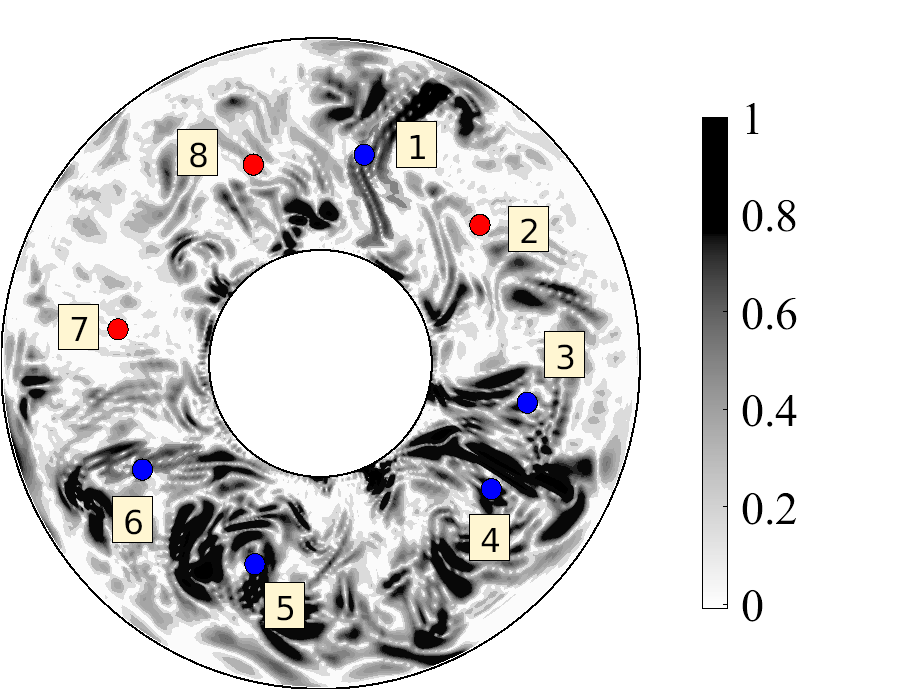}
		\includegraphics[width=0.32\linewidth]{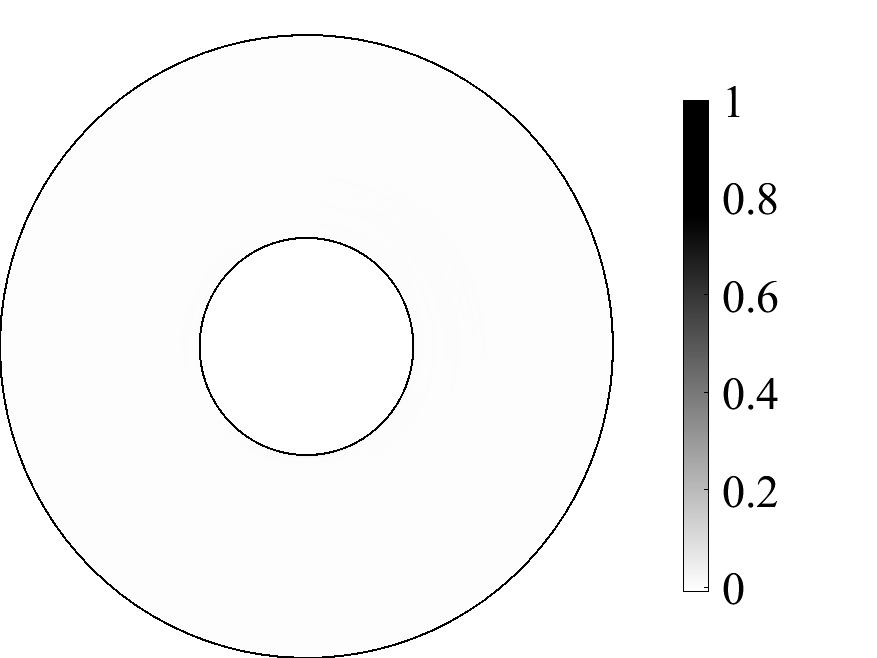}
		\hspace*{-0.75cm} \raisebox{-0.4cm}
		{\includegraphics[width=0.32\linewidth]{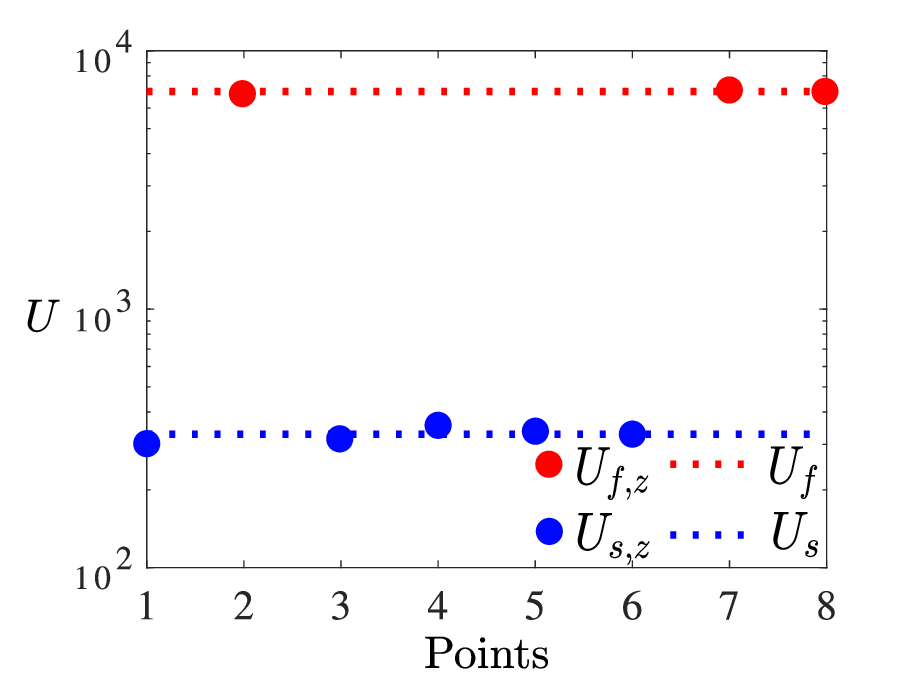}}\\
		\caption{Greyscale plots on a horizontal section at $z = 0.2$ 
			below the equatorial plane at point \lq B' 
			in figure \ref{tiltfdip}(c), 
			showing the ratio of the magnitudes of (a) Coriolis, 
			(b) Lorentz, (c) buoyancy, (d) non-dipolar Lorentz, and (e) 
			dipolar Lorentz force terms in the $z$-vorticity equation to 
			the magnitude of the largest force among them. 
			(f) The estimated and measured group velocities for 
			the blue and red points shown for figure (a) and (d). 
			Here, $U_{f,z}$ and $U_{s,z}$ denote the 
			measured axial group velocities of the fast and 
			slow MAC waves, respectively, and $U_{f}$ and 
			$U_{s}$ give their mean estimated 
			axial group velocities.
			The other dynamo parameters are 
			$Ra^C=3000,Ra^T=220, 
			E = 6 \times 10^{-5}$, and $Pm=Sc=5,Pr=0.5$. }
		\label{forces}
	\end{figure}
	%
	%
	%\begin{figure}
	%	\centering
	%	\hspace{-1.75 in}	(a)  \hspace{1.5 in} (b)  \hspace{1.5 in} (c) \\
	%	\includegraphics[width=0.32\linewidth]{coriolishydro.eps}
	%	\includegraphics[width=0.32\linewidth]{buoyancyhydro.eps}
	%	\raisebox{-0.4cm}
	%	{\includegraphics[width=0.32\linewidth]{pointshydro.eps}}
	%	\caption{Greyscale plots on a horizontal section 
		%	at $z = 0.2$ below the equatorial plane for 
		%	a hydrodynamic simulation, showing the ratio 
		%	of the magnitudes of (a) Coriolis and 
		%	(b) buoyancy force terms in the 
		%	$z$-vorticity equation to the magnitude of 
		%	the largest force among them. 
		%	(c) The estimated and measured group 
		%	velocities for the red points 
		%	in figure (a). Here, $U_{f,z}$ 
		%   denotes the measured axial 
		%	group velocities of the fast MAC waves and 
		%	$U_{f}$ denotes 
		%	their mean estimated axial group velocity.
		%	The other dynamo parameters are $Ra^C=3000,
		%	Ra^T=220, E = 6 \times 10^{-5}$, and 
		%	$Pm=Sc=5,Pr=0.5$. }
	%	\label{forceshydro}
	%\end{figure}
	%
	%
	%
	\begin{figure}
		\centering
		{\large	Dynamo simulation  }\\\vspace*{0.5cm}
		{Time \lq B' for $Ra^T = 220$, $Ra^C = 3000$}\\
		\hspace{-2.5 in}	(a)  \hspace{2.5 in} (b) \\
		\includegraphics[width=0.48\linewidth]{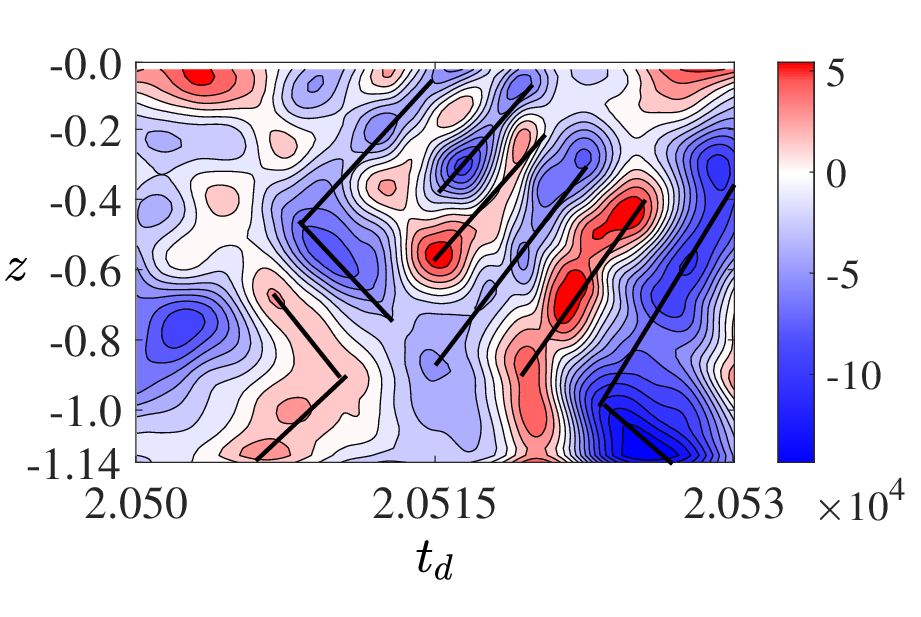}
		\includegraphics[width=0.48\linewidth]{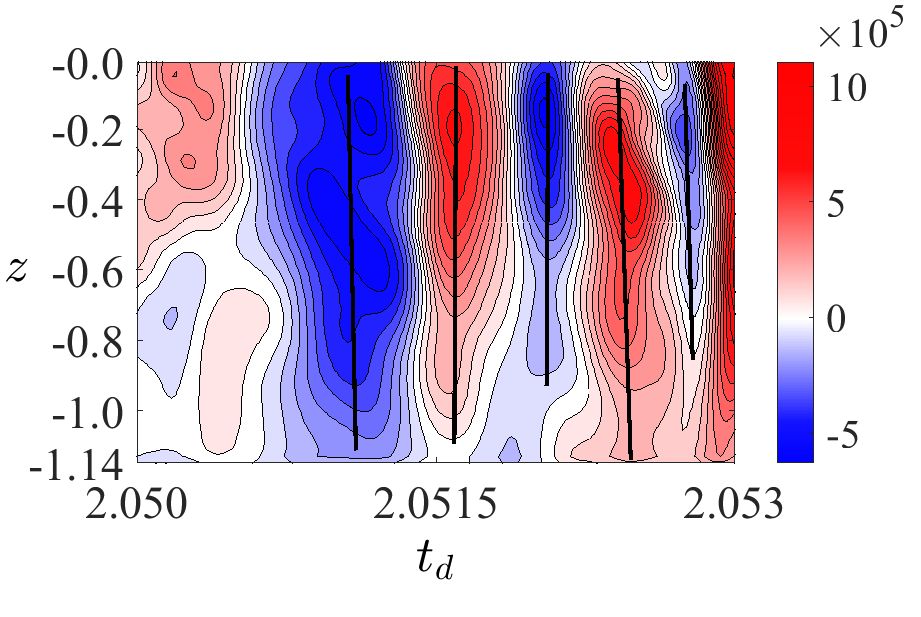}\\
		{\large	Hydrodynamic simulation}\\
		\hspace{-2.5 in}	(c)  \hspace{2.5 in} (d) \\
		\includegraphics[width=0.48\linewidth]{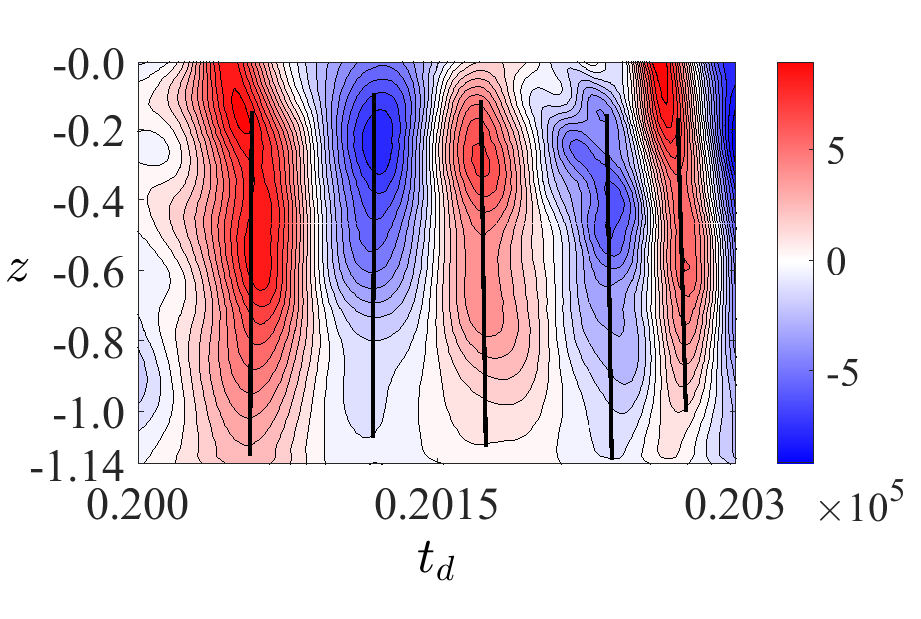}
		\includegraphics[width=0.48\linewidth]{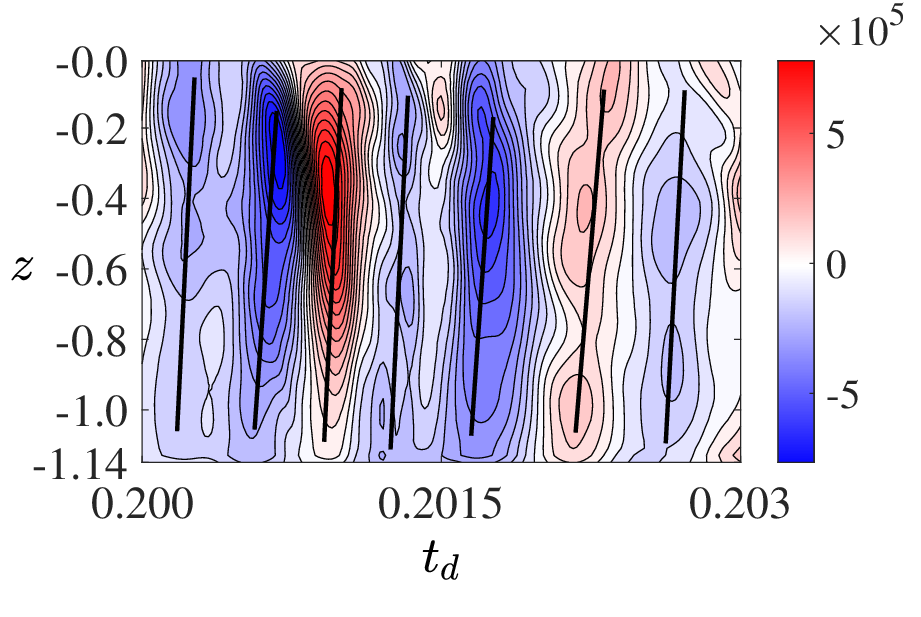}\\
		\caption{Contour plots of $\partial u_z / \partial t$ 
			at a cylindrical 
			radius of $s = 1$ for $l\le l_E$ are presented for 
			time \lq B' in the dynamo 
			at points (a) \lq 1' and (b) \lq 2' in 
			figure~\ref{forces}.  Similar contour
			plots for the hydrodynamic simulation
			at points (c) \lq 1' and 
			(d) \lq 2' in figure S2, Supporting Information.		
			The parallel black lines indicate the primary 
			wave travel direction, 
			with their slopes giving the measured group velocity $U_{g,z}$. 
			The other dynamo parameters are $Ra^C = 3000$, $Ra^T = 220$, 
			$E = 6 \times 10^{-5}$, $Pm = Sc = 5$, and $Pr = 0.5$.}
		\label{cga}
	\end{figure}
	%
	%
	%\begin{figure}
	%	\centering
	%	\hspace{-0.8 in}	(a) \hspace{0.35 in} Time `A' \hspace{0.8 in}
	%	(b) \hspace{0.35 in} Time `B' \hspace{0.8 in}
	%	(c) \hspace{0.35 in}Time `B' \\
	%	\includegraphics[width=0.30\linewidth]{cgta.eps}
	%	\includegraphics[width=0.33\linewidth]{cgs1.eps}
	%	\includegraphics[width=0.33\linewidth]{cgf3.eps}\\
	%	\caption{Contour plots of $\partial u_z / \partial t$ at a cylindrical 
		%		radius of $s = 1$ for $l\le l_E$ are presented for (a) time `A'; 
		%		for time `B' at points (b) `1' and (c) `2' in 
		%		Figure~\ref{forces}.		
		%		The parallel black lines indicate the primary wave travel direction, 
		%		with their slopes denoting the group velocity $U_{g,z}$. 
		%		Panels (a) and (c) show the propagation of fast MAC waves, 
		%		whereas panel (b) shows the propagation of slow MAC waves. 
		%		The other dynamo parameters are $Ra^C = 3000$, $Ra^T = 220$, 
		%		$E = 6 \times 10^{-5}$, $Pm = Sc = 5$, and $Pr = 0.5$. }
	%	\label{cg}
	%\end{figure}
	In Section \ref{linear}, it was shown that 
	an isolated density perturbation in an 
	unstably stratified fluid subject to 
	background rotation and a magnetic field 
	produces MAC waves. The frequencies
	of these waves depend 
	on the linear 
	inertial wave frequency $\omega_C$, the 
	Alfv\'en wave frequency $\omega_M$ and 
	the buoyancy frequencies $\omega_{A}^C$ 
	and $\omega_{A}^T$ (see equations 
	\eqref{l12approx}--\eqref{l34approx}). 
	For the simulation in 
	figure \ref{tiltfdip}(c),
	the fundamental and slow MAC frequencies,
	scaled by $\eta/L^2$, evolve as 
	shown in figure \ref{freqtseries}(a). 
	The frequencies are based on the mean
	wavenumbers $\bar{m}$, $\bar{k}_s$ and
	$\bar{k}_z$ given in Table S1,
	Supporting Information.
	As described earlier, the simulation began 
	with purely compositional 
	convection, which exhibited multiple polarity reversals. 
	At time marked \lq A’, a finite thermal 
	buoyancy was introduced, 
	contributing approximately 20\% of the total 
	convective power. A dipole solution with 
	the frequency inequality 
	$|\omega_C| > |\omega_M| > |\omega_A|$ ensues. 
	Figures \ref{freqtseries}(b, c) show that 
	$|\omega_M^2| > |\omega_A^2|$ holds over
	several points at time \lq C', indicating
	slow MAC wave activity. The slow MAC waves
	of frequency $\omega_s$ emerge near time 
	\lq B' when the field is still multipolar. At this time, 
	the Lorentz force is dominated by non-dipolar 
	magnetic field 
	(figures \ref{forces} d, e).
	Figure \ref{cga} (a, b) shows the wave motions at time 
	\lq B' when both fast and slow 
	MAC waves are found. The slow waves are 
	generated in regions of concentrated magnetic flux, 
	where balances between Lorentz, Coriolis, 
	and buoyancy forces are established. 
	The fast MAC waves exist in 
	areas where the magnetic field is 
	comparatively weak. The axial group velocities 
	are also estimated by taking the derivative
	of the respective frequency with respect to
	the $z$ wavenumber, $k_z$ \cite{aditya2022}, and given in
	figure \ref{forces}(f).  
	The helicity associated with slow MAC waves 
	generated by non-dipolar magnetic field 
	plays a crucial role in the subsequent formation 
	of the axial dipole \cite{aditya2022,jfm24}.  
	
	A purely hydrodynamic simulation 
	with identical control parameters, where the 
	Lorentz force is absent, remains in a multipolar 
	state \cite{prf18}, highlighting the essential 
	role of the slow MAC waves in dipole 
	formation. In this hydrodynamic dynamo,
	the ageostrophic Coriolis force balances the
	buoyancy force (figure S2, Supporting Information),
	and only fast MAC waves 
	are observed (figure \ref{cga} c, d).
	\begin{figure}
		\centering
		\hspace{-2.25 in}(a)  \hspace{2.25 in} (b) \\
		\includegraphics[width=0.45\linewidth]{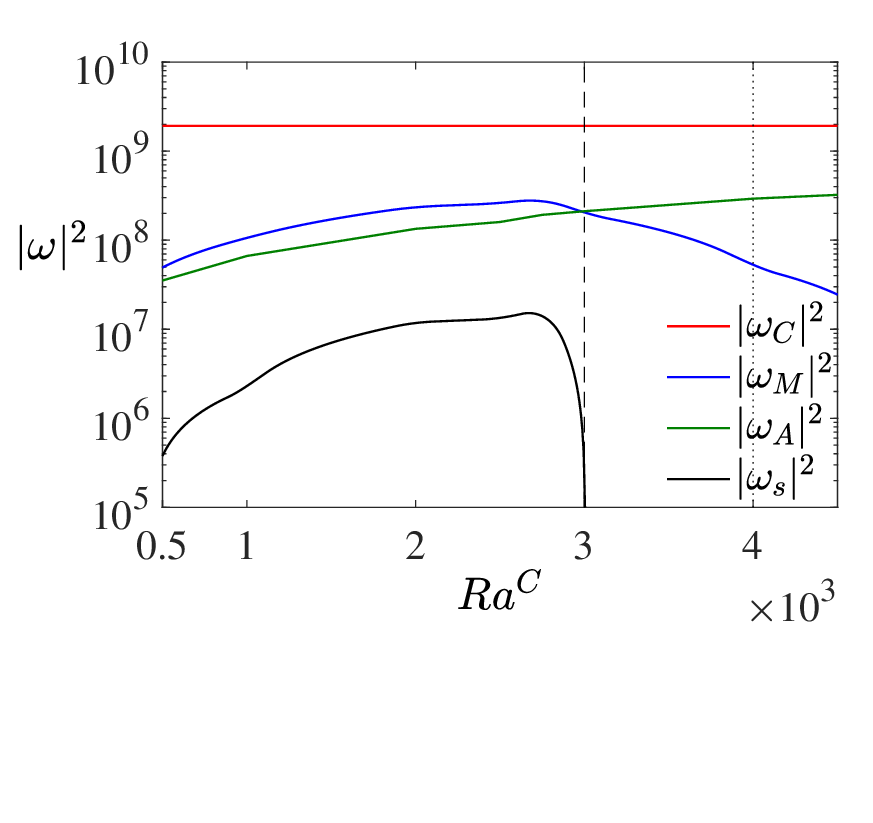}
		\includegraphics[width=0.45\linewidth]{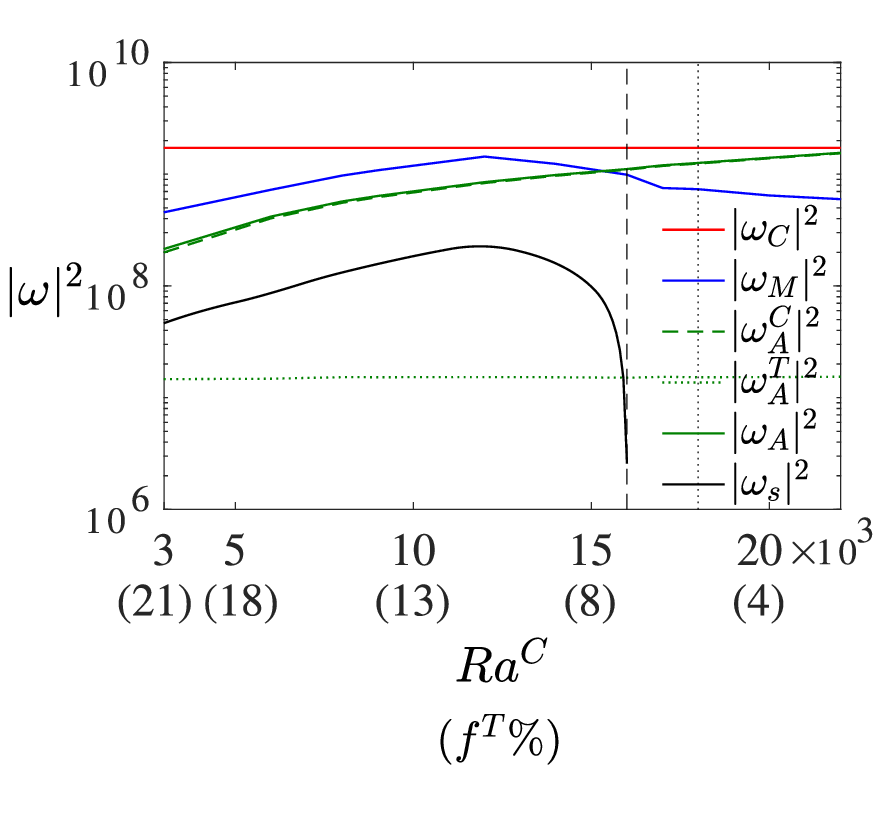}\\\vspace*{-1cm}
		\hspace{-2.25 in}(c)  \hspace{2.25 in} (d) \\
		\includegraphics[width=0.45\linewidth]{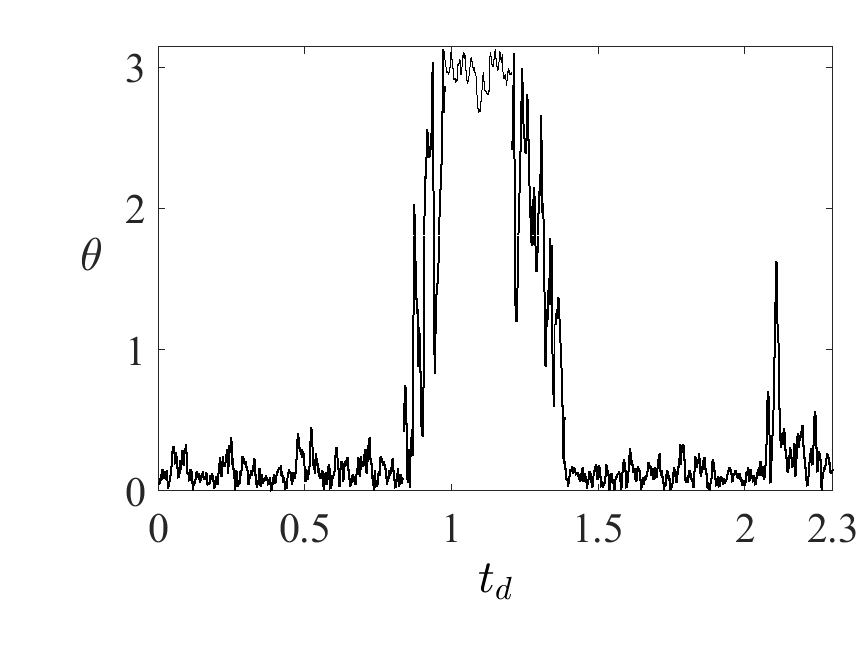}
		\includegraphics[width=0.45\linewidth]{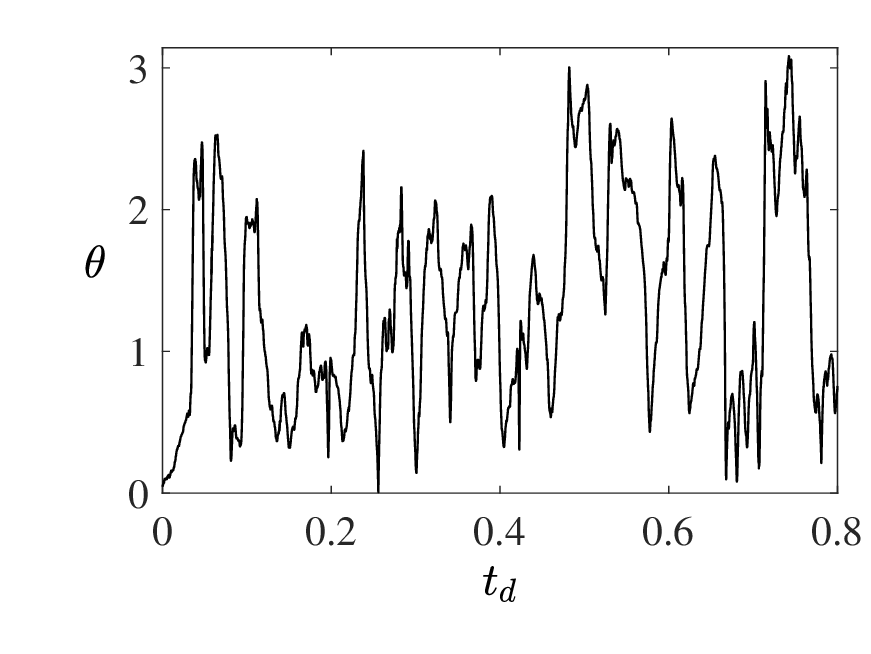}\\
		\caption{(a--b) Absolute values of the squared fundamental 
			frequencies in the saturated state as a function of 
			$Ra^C$ for (a) one-component and (b) two-component 
			convection. In (b), the thermal Rayleigh number 
			is fixed, contributing about 20\% of the 
			total convective power with the one-component 
			case at $Ra^C = 3000$. The dashed vertical line in 
			(b) shows where the slow MAC wave frequency $\omega_s$ 
			goes to zero, when $|\omega_{M}| \approx |\omega_A|$, 
			marking the onset of reversals and the dotted line marks 
			the onset of the multipolar regime. (c--d) Evolution of 
			the dipole colatitude as a function of the magnetic diffusion 
			time for (c) $Ra^T = 220$, $Ra^C = 16000$ (dashed line in 
			(b), showing reversals) and (d) $Ra^T = 220$, $Ra^C = 18000$ 
			(dotted line in (b), showing the multipolar state). 
			The dynamo parameters are $E = 6 \times 10^{-5}$, 
			$Pm = Sc = 5$, $Pr = 0.5$.}	
		\label{freqsddd}
	\end{figure}
	\subsection{Extension of the axial dipole regime
		in two-component convection}
	The addition of thermal buoyancy to a compositionally
	driven reversing dynamo produces a stable axial dipole 
	in two-component convection. 
	To make the dynamo reverse polarity again, 
	the compositional Rayleigh number must be increased 
	until the magnitudes of the resultant buoyancy frequency 
	$\omega_A$ and the the Alfv\'en frequency $\omega_M$
	match again so that
	the slow MAC wave frequency $\omega_s$ goes to zero.
	This provides the central result of this study, that
	the range of compositional
	buoyancy that admits the axial dipole is much wider
	in two-component convection relative to one-component
	convection. 
	Figure~\ref{freqsddd} 
	shows the variation of the fundamental 
	frequencies with the compositional Rayleigh number $Ra^C$
	for (a) purely compositional convection 
	and (b) two-component thermochemical convection, 
	in the saturated states of the dynamos. 
	For purely compositional convection, 
	a Rayleigh number of 
	$Ra^C = 3000$ is sufficient to trigger polarity
	reversals. In two-component convection,  
	reversals via the suppression of
	slow MAC waves occur at $Ra^C = 16000$, corresponding
	to a thermal power ratio of $\approx$ 7.5 \%. 
	The dashed vertical lines in figures \ref{freqsddd} 
	(a) and (b) 
	mark the onset of the reversal regime 
	and the dotted lines indicate the transition 
	into the multipolar regime. Figures~\ref{freqsddd}(c) 
	and (d) show the evolution of the dipole 
	colatitude for the two-component cases with
	$Ra^T = 220$, $Ra^C = 16000$ (dashed line in 
	figure \ref{freqsddd}b) and $Ra^T = 220$, 
	$Ra^C = 18000$ (dotted line in figure \ref{freqsddd}b), 
	respectively. Evidently, the former represents a 
	reversing regime whereas the latter 
	represents a multipolar regime.
	For purely compositional convection
	at $E=1.2 \times 10^{-5}$, 
	$Ra^C = 12000$ triggers polarity
	reversals. Two-component convection
	with $Ra^C=20000$ and $Ra^T=500$
	produces a non-reversing dipole (Table \ref{tablepara}).  
	For a fixed $Ra^T$, reversals 
	are anticipated at $Ra^C \sim 60000$,
	a computationally demanding parameter space.
	
	The extension of the range of the axial dipole in
	the nonlinear dynamo simulations may be compared with
	that predicted in linear magnetoconvection (Section
	\ref{macwtcc}). From the two- and one-component states
	on the threshold of the polarity transition, the ratio of
	Rayleigh numbers is $\approx$ 5.3 while the respective ratio
	of the squares of the peak field intensities is
	$\approx 105/20 \approx 5.25$ (see Table \ref{tablepara}).
	For a thermal power ratio of $\approx$ 7.5\%, the Rayleigh
	number ratio and the ratio of
	squares of the field intensities are both $\approx$ 4 as suggested
	by the linear relation between the two (figure 
	\ref{eklinear}b), which
	indicates that the analysis of quasi-linear MAC wave motions
	in magnetoconvection subject to the conditions
	\eqref{cons1} and \eqref{cons2}
	provides a fair insight into the range of the axial dipole
	in rapidly rotating low-inertia dynamos. Although the linear
	model predicts a polarity transition for any power ratio $f^T$,
	for $f^T >$ 8\% the self-consistent dynamo with homogeneous boundary heat flux
	generates a sufficiently intense field 
	that ensures $|\omega_M| > |\omega_A|$, or in
	other words, dipole dominance.
	To induce a polarity transition for $f^T >$ 8\%, an additional
	source of buoyancy must be included, originating from
	laterally varying outer boundary heat flux (Section
	\ref{conclusion}). 
	\section{Discussion}
	\label{conclusion}
	
	\begin{figure}
		\centering
		\hspace{-2.25 in}	(a)  \hspace{2.25 in} (b) \\
		\includegraphics[width=0.44\linewidth]{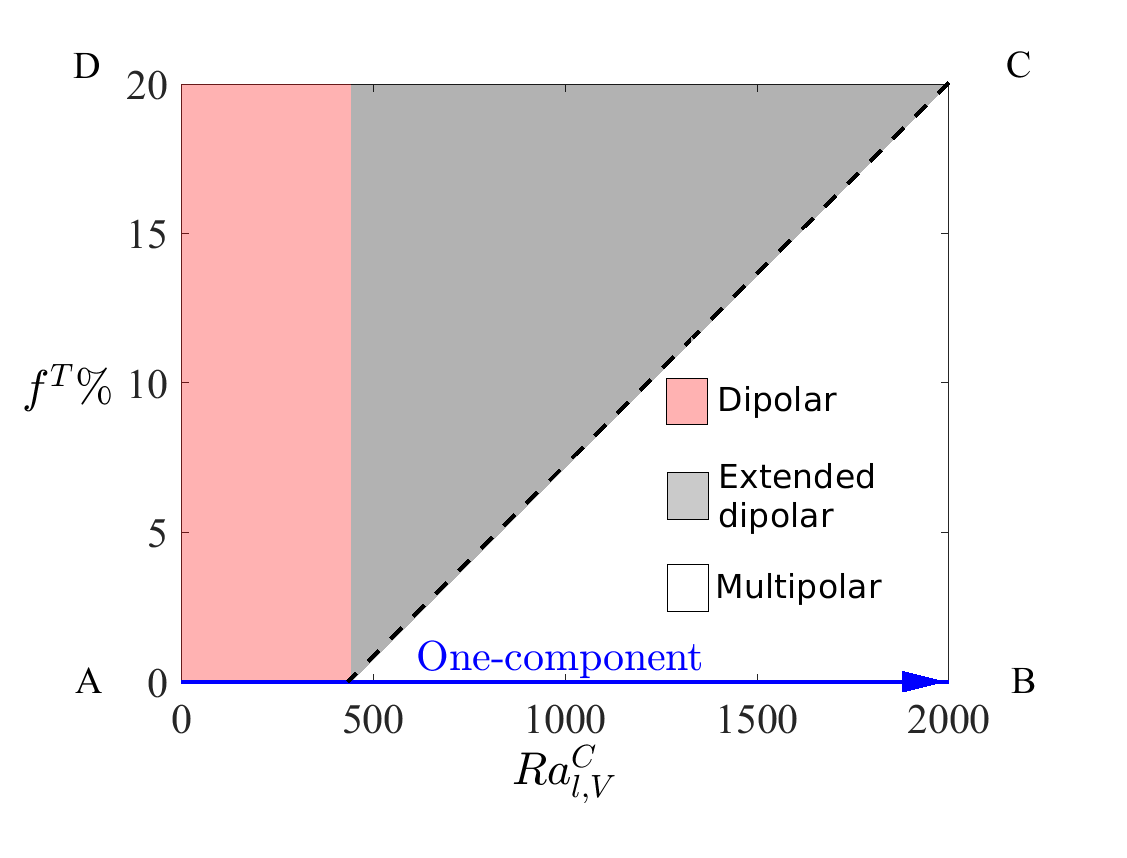}
		\includegraphics[width=0.45\linewidth]{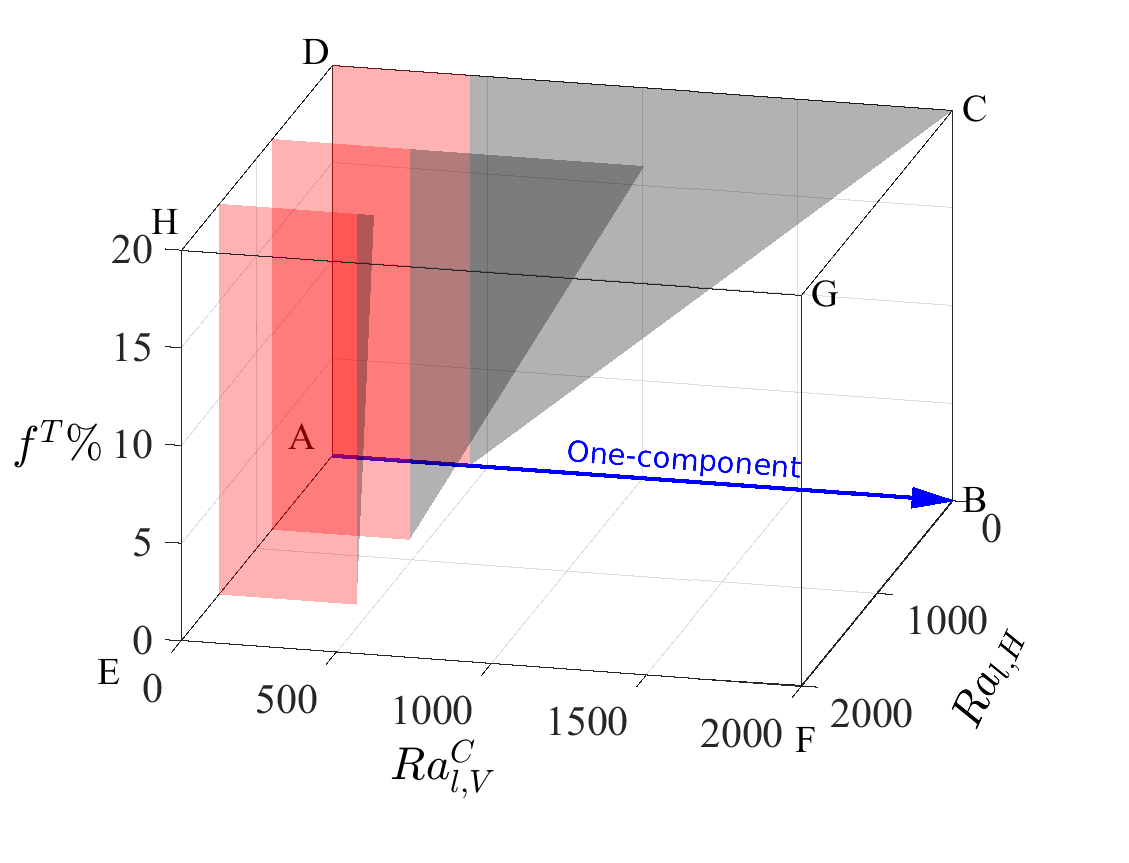}
		\caption{(a) Schematic diagram illustrating the 
			extent of slow MAC wave activity, 
			which can also be interpreted 
			as the range of stability 
			of the axial dipole field. 
			The slow MAC 
			waves are always present in the red region,
			and the axial 
			dipolar field remains stable here. 
			The grey region denotes an extended 
			regime in which the presence 
			of slow MAC waves, and hence that of the 
			the axial dipole, 
			depends on the available thermal power. 
			The white region represents the 
			absence of slow MAC waves, 
			giving reversing 
			and multipolar dynamos. 
			The dashed line separating the 
			grey and white regions 
			marks the transition between the presence 
			and absence of slow MAC waves, 
			corresponding to the 
			dipole--multipole transition. 
			The blue horizontal line indicates 
			the evolution of the 
			dynamo in one-component convection as the 
			compositional forcing is increased. 
			(b) Schematic of the axial dipole stability, 
			of which panel (a) is a subset. 
			Here, the 
			existence of the axial dipole 
			depends on both vertical buoyancy 
			contributions from 
			thermal and compositional parts, 
			as well as on the horizontal 
			(lateral) buoyancy arising from 
			the heat flux heterogeneity at the CMB.
			The diagrams are based on linear magnetoconvection
			calculations at $E_\eta = 1 \times 10^{-5}$.
			Here, $Ra_{\ell,V}^C$ and $Ra_{\ell,H}$ are
			the vertical Rayleigh number of composition
			and the horizontal Rayleigh number, both based on 
			the scale of a density disturbance.}
		\label{schematic2}
	\end{figure}
	
	This study focuses on two-component convection
	in Earth-like low-inertia dynamos, where 
	nonlinear inertia is small compared with the Coriolis force 
	not only at the planetary scale but also at the characteristic 
	length scale of convection. In the models considered here, 
	compositional buoyancy provides the primary 
	driving force. However, supercritical
	states of buoyancy flux $> O(10^2)$ times 
	the critical buoyancy flux for onset of convection do not
	support an axial dipole field. In addition, the peak
	azimuthal speed of the polar vortices scaled up to Earth's core 
	are markedly
	weaker than the observed
	peak drift (Table \ref{tablepara}). 
	Since much stronger compositional buoyancy fluxes would
	only produce multipolar fields, 
	compositional buoyancy must be paired with relatively
	weak thermal buoyancy to retain the axial dipole character of
	the field as well as the observed polar circulation.
	The numerical experiments in this study suggest that
	$Ra^C = O(10^3) \times$ its onset value and $Ra^T =O(1)
	\times$ its onset value produce dipole-dominated
	dynamos with Earth-like polar drift rates and
	thermal power ratios in the range $f^T=$10--25\%.
	
	%The addition of thermal buoyancy 
	%introduces two distinct and competing effects. First, it increases 
	%the total buoyancy through the addition of a second buoyant component. 
	%Second, and more importantly, it enhances magnetic field generation, 
	%leading to an increase in the effective Alfv\'en wave frequency 
	%$\omega_M$. As a result, $\omega_M$ can exceed the resultant 
	%buoyancy frequency $\omega_A$, thereby restoring the conditions 
	%required for the existence of slow MAC waves and the maintenance 
	%of a stable axial dipolar magnetic field.
	
	Figure \ref{schematic2}(a) illustrates
	the parameter space in which slow MAC 
	waves are supported, which is essentially
	the stability 
	domain of the axial dipole. This diagram is based on
	the linear magnetoconvection model 
	in Section \ref{linear}. The red region denotes the parameter 
	range in which slow MAC waves, and hence a dipole field, 
	are supported in purely compositional convection. 
	As the compositional Rayleigh number based on
	the scale of a density disturbance, $Ra_{\ell,V}^C$, 
	progressively increases, the 
	slow MAC waves are suppressed. 
	When thermal buoyancy is added, however, the
	range of $Ra_{\ell,V}^C$ over which slow MAC waves exist widens. 
	This extended window, shown 
	in grey, depends on the 
	available thermal power. For thermal power
	ratios $f^T=$ 10--25\%, the nonlinear dynamo solutions are
	non-reversing dipoles since $|\omega_M|>|\omega_A|$. 
	The moderately enhanced core heat fluxes during the
	Cretaceous Normal Superchron suggested by mantle
	dynamics models \cite{zhang2011heat,olson2013controls}
	may enhance $f^T$.
	Numerical experiments with higher
	$f^T$, obtained by increasing $Ra^T$ at a fixed
	$Ra^C$ (Table \ref{tablepara}), again produce non-reversing
	dipoles. These results indicate that Earth's core need not
	operate near the boundary of polarity transitions, marked
	by the dashed line that separates the grey and white regions
	in figure \ref{schematic2}(a).
	
	%%These results have important geological implications, discussed below.
	As noted thus far, in the low-inertia limit, a two-component dynamo with
	homogeneous outer boundary heat flux produces 
	a dipole-dominated field for thermal
	power ratios $f^T>$10\%. Polarity
	transitions are possible for $f^T<8\%$, 
	a scenario not considered likely for Earth,   
	which has passed through extended periods without
	polarity reversals.
	To explain how field
	excursions and occasional reversals
	occur in Earth, one must introduce an additional
	source of buoyancy, the origin of which lies in the
	laterally varying heat flux in the lowermost mantle. 
	The lateral (horizontal)
	buoyancy, measured by its Rayleigh number $Ra_{\ell,H}$,
	does not alter the thermal power ratio, and its
	contribution to dynamo field generation is negligible
	in the strongly driven core conditions considered here. 
	That said, in an unstably stratified fluid, 
	the vertical and horizontal buoyancy frequencies 
	complement each other such that the resultant 
	buoyancy frequency approximately matches 
	the Alfv\'en wave frequency in order to 
	suppress slow MAC waves and thereby induce 
	magnetic polarity reversals \cite{arxiv25}.
	In particular, the horizontal buoyancy must consist
	of an equatorially anti-symmetric part to induce
	reversals. Figure \ref{schematic2}(b), of
	which figure (a) is a subset, shows
	how a dynamo that resides deep within the dipolar
	regime can proceed to a reversing state by a progressive
	substantial increase
	in $Ra_{\ell,H}$ over geological time. Small
	values of $Ra_{\ell,H}$ may correspond to either
	a lower lateral variation in heat flux at the core
	surface or a decrease in the equatorially anti-symmetric part
	of the variation relative to the symmetric part, which
	would result in the periods without polarity reversals.
	
	%%%%%%%%%%%%%%%%%%%%%%%%%%%%%%%%%%%%%%%%%%%%%%%%%%%%
	\clearpage
	\newpage
	
	\appendix
	
	\section{Calculation of the initial wavenumber in linear magnetoconvection}
	\label{lmc}
	%%%
	The Fourier transform of the initial condition \eqref{pert} is,
	\begin{equation}
		\hat{\Theta}_0 = \pi^{3/2} \delta^3 
		\exp{\left(-\frac{k^2\delta^2}{4} \right)},
	\end{equation}
	where $k=\sqrt{k_x^2+k_y^2+k_z^2}$. 
	
	The initial wavenumber is defined by,
	\begin{equation}
		k_0=\left[\dfrac{\displaystyle \int_{\!\!-\infty}^{\infty} 
			\displaystyle \int_{\!\!-\infty}^{\infty}
			\displaystyle \int_{\!\!-\infty}^{\infty} 
			k^2 |\hat{\Theta}_0|^2
			\,\mbox{d}k_x\,\mbox{d}k_y\,
			\mbox{d}k_z}{\displaystyle \int_{\!\!-\infty}^{\infty}
			\displaystyle\int_{\!\!-\infty}^{\infty}
			\displaystyle \int_{\!\!-\infty}^{\infty}
			|\hat{\Theta}_0|^2\, \mbox{d}k_x\, 
			\mbox{d}k_y\, \mbox{d}k_z}\right]^{1/2}.
	\end{equation}
	Letting $k_x=k \sin\phi \cos\theta$, $k_y= 
	k \sin\phi \sin\theta$, $k_z = k \cos\phi$, 
	we obtain
	\begin{eqnarray}
		k_0 &=& \left[\frac{\displaystyle \int_{0}^{2 \pi} 
			\displaystyle \int_{0}^{\pi}
			\displaystyle \int_{0}^{\infty} k^2 
			|\pi^{3/2}\delta^3\exp{\left(-k^2\delta^2/4 \right)}|^2
			k^2 \sin\phi \,\mbox{d}k\, 
			\mbox{d}\theta\, \mbox{d}\phi}{\displaystyle \int_{0}^{2 \pi}
			\displaystyle \int_{0}^{\pi} \displaystyle 
			\int_{0}^{\infty}|\pi^{3/2}\delta^3  
			\exp{\left(-k^2\delta^2/4 \right)}|^2k^2 
			\sin \phi\, \mbox{d}k\, \mbox{d}\theta\, 
			\mbox{d}\phi}\right]^{1/2}, \\
		&=& \sqrt{3}/\delta,
	\end{eqnarray}
	
	on evaluation of the integrals. Considering equal
	values for the initial wavenumbers, we obtain
	$\{k_{x,0},k_{y,0},k_{z,0}\} = k_0/\sqrt{3}=1/\delta$.
	%%
	
	%
	%\bibliographystyle{apalike}
	%\bibliography{library1}
	%\end{document}

	\acknowledgments
	This study was supported in part by Research Grant 
	MoE-STARS/STARS-1/504 under
	Scheme for Transformational 
	and Advanced Research in Sciences 
	awarded by the Ministry of  
	Education (India) and in part 
	by Research grant CRG/2021/002486 
	awarded by the Science and 
	Engineering Research Board (India). 
	The computations were performed on Param Pravega, 
	the  supercomputer at the 
	Indian Institute of Science, Bangalore.

	%%%%%%%%%%%%%%%%%%%%%%%%%%%%%%%%%%%%%%%%%%%%%%%
	% REFERENCES and BIBLIOGRAPHY
	%
	\bibliography{library1}

@article{jones2015thermal,
  title={Thermal and compositional convection in the outer core},
  author={Jones, C. A. and Schubert, G},
  journal={Treatise in Geophysics, Core Dynamics},
  volume={8},
  pages={131--185},
  year={2015}
}

@article{sreenivasan2017confinement,
  title={Confinement of rotating convection by a laterally varying magnetic field},
  author={Sreenivasan, B. and Gopinath, V.},
  journal={J. Fluid Mech.},
  volume={822},
  pages={590--616},
  year={2017},
  publisher={Cambridge University Press}
}

@article{jfm21,
  title={Evolution of forced magnetohydrodynamic waves in a stratified fluid},
  author={Sreenivasan, B. and Maurya, G.},
  journal={J. Fluid Mech.},
  volume={922},
  year={2021},
  publisher={Cambridge University Press}
}

@article{manglik2010dynamo,
  title={A dynamo model with double diffusive convection for Mercury's core},
  author={Manglik, A. and Wicht, J. and Christensen, U. R.},
  journal={Earth Planet. Sci. Lett.},
  volume={289},
  number={3-4},
  pages={619--628},
  year={2010},
  publisher={Elsevier}
}

@incollection{07bussechapter,
  title={Dynamics of rotating fluids},
  author={Busse, F. and Dormy, E. and Simitev, R. and Soward, A.},
  booktitle={Mathematical Aspects of Natural Dynamos},
  volume={13},
  pages={165--168},
  year={2007},
editor = {Dormy, E. and Soward, A. M.},
series = {The Fluid Mechanics of Astrophysics and Geophysics},
  publisher={CRC Press}
}

@article{jfm24,
  title={Self-similarity of the dipole--multipole 
  transition in rapidly rotating dynamos},
  author={Majumder, D. and Sreenivasan, B. and Maurya, G.},
  journal={J. Fluid Mech.},
  volume={980},
  pages={A30},
  year={2024},
  publisher={Cambridge University Press}
  }

@article{brag1967,
  title={Magnetic waves in the {E}arth's core},
  author={Braginsky, S. I.},
  journal={Geomagn. Aeron.},
  volume={7},
  pages={851--859},
  year={1967}
}

@article{braginsky1995equations,
  title={Equations governing convection in {E}arth's core and the geodynamo},
  author={Braginsky, S. I. and Roberts, P. H.},
  journal={Geophys. Astrophys. Fluid Dyn.},
  volume={79},
  number={1-4},
  pages={1--97},
  year={1995},
  publisher={Taylor \& Francis}
}

@article{poirier1994light,
  title={Light elements in the {E}arth's outer core: {A} critical review},
  author={Poirier, J.-P.},
  journal={Phys. Earth Planet. Int. 85,},
  volume={85},
  number={3-4},
  pages={319--337},
  year={1994},
  publisher={Elsevier}
}

@article{pozzo2012thermal,
  title={Thermal and electrical conductivity of iron at {E}arth's core conditions},
  author={Pozzo, M. and Davies, C. and Gubbins, D. and Alfe, D.},
  journal={Nature},
  volume={485},
  number={7398},
  pages={355--358},
  year={2012},
  publisher={Nature Publishing Group UK London}
}

@article{lister1995strength,
  title={The strength and efficiency of thermal 
  and compositional convection in the geodynamo},
  author={Lister, J. R. and Buffett, B. A.},
  journal={Phys. Earth Planet. Inter.},
  volume={91},
  number={1-3},
  pages={17--30},
  year={1995},
  publisher={Elsevier}
}

@article{loper2003buoyancy,
  title={Buoyancy-driven perturbations in a rapidly rotating,
   electrically conducting fluid: part {I}--flow and magnetic field},
  author={Loper, D. E. and Chulliat, A. and Shimizu, H.},
  journal={Geophys. Astrophys. Fluid Dyn.},
  volume={97},
  number={6},
  pages={429--469},
  year={2003},
  publisher={Taylor \& Francis}
}

@ARTICLE{07willis,
        author = {A. P. Willis and B. Sreenivasan and
D. Gubbins},
        title = {Thermal core-mantle interaction: {E}xploring regimes
for \lq locked' dynamo action},
        journal = {Phys. Earth Planet. Inter.},
        volume = {165},
        pages = {83--92},
        year = {2007},
}

@article{chraub2006,
  title={Scaling properties of convection-driven dynamos 
in rotating spherical shells and application to planetary magnetic fields},
  author={Christensen, U. R. and Aubert, J.},
  journal={Geophys. J. Int.},
  volume={166},
  number={1},
  pages={97--114},
  year={2006},
  publisher={Blackwell Publishing Ltd Oxford, UK}
}

@article{takahashi2014double,
  title={Double diffusive convection in the {E}arth's core and the morphology of the geomagnetic field},
  author={Takahashi, F.},
  journal = {Phys. Earth Planet. Inter.},
  volume={226},
  pages={83--87},
  year={2014},
  publisher={Elsevier}
}

@book{glatz2013,
  title={Introduction to Modeling Convection in Planets
  and Stars: Magnetic Field, Density Stratification, Rotation},
  author={Glatzmaier, G. A.},
  year={2013},
  publisher={Princeton University Press},
}

@article{aditya2022,
  title = {The role of slow magnetostrophic waves 
in the formation of the axial dipole
in planetary dynamos},
  author = {A. Varma and B. Sreenivasan},
  journal = {Phys. Earth Planet. Inter.},
  volume = {333},
  pages = {106944},
  year = {2022},
}

@article{takahashi2019mercury,
  title={Mercury’s anomalous magnetic field caused by a symmetry-breaking self-regulating dynamo},
  author={Takahashi, F. and Shimizu, H. and Tsunakawa, H.},
  journal={Nat. Commun.},
  volume={10},
  number={1},
  pages={208},
  year={2019},
  publisher={Nature Publishing Group UK London}
}

@article{glatzmaier1996anelastic,
  title={An anelastic evolutionary geodynamo simulation driven by compositional and thermal convection},
  author={Glatzmaier, G. A. and Roberts, P. H.},
  journal={Physica D.},
  volume={97},
  number={1-3},
  pages={81--94},
  year={1996},
  publisher={Elsevier}
}

@article{tassin2021geomagnetic,
  title={Geomagnetic semblance and dipolar--multipolar transition 
  in top-heavy double-diffusive geodynamo models},
  author={Tassin, T. and Gastine, T. and Fournier, A.},
  journal={Geophys. J. Int.},
  volume={226},
  number={3},
  pages={1897--1919},
  year={2021},
  publisher={Oxford University Press}
}

@article{zhang2011heat,
  title={Heat fluxes at the {E}arth's surface and core--mantle boundary 
  since {P}angea formation and their implications for the geomagnetic superchrons},
  author={Zhang, N. and Zhong, S.},
  journal={Earth Planet. Sci. Lett.},
  volume={306},
  number={3-4},
  pages={205--216},
  year={2011},
  publisher={Elsevier}
}

@article{olson2013controls,
  title={Controls on geomagnetic reversals and core 
  evolution by mantle convection in the {P}hanerozoic},
  author={Olson, P. and Deguen, R. and Hinnov, L. A. and Zhong, S.},
  journal = {Phys. Earth Planet. Inter.},
  volume={214},
  pages={87--103},
  year={2013},
  publisher={Elsevier}
}

@article{prf18,
    author = "Sreenivasan, B. and Kar, S.",
    title = "Scale dependence of kinetic helicity and selection 
    of the axial dipole in rapidly rotating dynamos",
    journal = "Phys. Rev. Fluids",
    volume = "3",
    number = "9",
    pages = "093801",
    year = "2018",
    publisher = "APS"
}

@article{arxiv25,
  title={Polarity transitions induced by symmetry-breaking outer boundary heat flux in rapidly rotating dynamos},
  author={Majumder, D. and Sreenivasan, B.},
  journal={arXiv preprint arXiv:2505.02673},
  year={2025}
}

@article{fan2025dynamos,
  title={Dynamos driven by top-heavy double-diffusive convection in the strong-field regime},
  author={Fan, W. and Lin, Y.},
  journal={ J. Geophys. Res.: Planets},
  volume={130},
  number={8},
  pages={e2025JE008969},
  year={2025},
  publisher={Wiley Online Library}
}

@article{mather2021regimes,
  title={Regimes of thermo-compositional convection and related dynamos in rotating spherical shells},
  author={Mather, J. F. and Simitev, R. D.},
  journal={Geophys. Astrophys. Fluid Dyn.},
  volume={115},
  number={1},
  pages={61--84},
  year={2021},
  publisher={Taylor \& Francis}
}

@article{labrosse2001age,
  title={The age of the inner core},
  author={Labrosse, S. and Poirier, J.-P. and Le Mou{\"e}l, J.-L.},
  journal={Earth Planet. Sci. Lett.},
  volume={190},
  number={3-4},
  pages={111--123},
  year={2001},
  publisher={Elsevier}
}

@article{bono2019young,
  title={Young inner core inferred from {E}diacaran ultra-low geomagnetic field intensity},
  author={Bono, R. K. and Tarduno, J. A. and Nimmo, F. and Cottrell, R. D.},
  journal={Nat. Geosci.},
  volume={12},
  number={2},
  pages={143--147},
  year={2019},
  publisher={Nature Publishing Group UK London}
}

@book{cormier2021earth,
  title={Earth's core: geophysics of a planet's deepest interior},
  author={Cormier, V. F. and Bergman, M. I. and Olson, P.},
  year={2021},
  publisher={Elsevier}
}

@article{hulot2002,
  title={Small-scale structure of the geodynamo inferred from
 {O}ersted and {M}agsat satellite data},
  author={Hulot, G. and Eymin, C. and Langlais, B. and Mandea, M. and Olsen, N.},
  journal={Nature},
  volume={416},
  number={6881},
  pages={620--623},
  year={2002},
  publisher={Nature Publishing Group UK London}
}

@article{grl2005,
    author = "Sreenivasan, B. and Jones, C. A.",
    title = "Structure and dynamics of the polar vortex in the {E}arth's core",
    journal = "Geophys. Res. Lett.",
    volume = "32",
    number = "20",
    year = "2005",
    publisher = "Wiley Online Library"
}

@article{lin2025invariance,
  title={Invariance of dynamo action in an early-{E}arth model},
  author={Lin, Y. and Marti, P. and Jackson, A.},
  journal={Nature},
  volume={644},
  number={8075},
  pages={109--114},
  year={2025},
  publisher={Nature Publishing Group UK London}
}

\clearpage
\newpage
	
\renewcommand{\tablename}{Table}
\renewcommand{\thetable}{S\arabic{table}}

\setcounter{figure}{0}
\renewcommand{\figurename}{Figure}
\renewcommand{\thefigure}{S\arabic{figure}}
\pagestyle{plain} % Suppress headers and use only page numbers
\renewcommand{\thesection}{S\arabic{section}}

\vspace{0cm}

\begin{center}
	Supporting Information for
	
	\vspace{0.2cm}
	
	\textbf{The role of thermal buoyancy in stabilizing the axial
		dipole field in rotating two-component convective dynamos}
	
	\vspace{0.2cm}
	
	\begingroup
	\renewcommand\thefootnote{\fnsymbol{footnote}}
	Debarshi Majumder and Binod Sreenivasan\footnote{Corresponding 
		author: Binod Sreenivasan (bsreeni@iisc.ac.in)}\\
	\endgroup
	Centre for Earth Sciences, 
	Indian Institute of Science, Bangalore 560012, India
\end{center}

\noindent
\textbf{Contents of this file}
\begin{itemize}
	\item Figures S1 \& S2
	\item Table S1
\end{itemize}

\vspace{0.5cm}

\noindent
\textbf{Introduction}

\vspace{0.3cm}

Figure S1 shows the induced axial magnetic energy of the fast and 
slow MAC waves in the linear magnetoconvection model (Section 2.3). 

Figure S2 shows the analysis of the forces 
and the wave group 
velocities in a hydrodynamic dynamo simulation (Section 3.2). 

Table S1 contains the spatial resolutions of the nonlinear dynamo 
simulations and the calculated wavenumbers (Section 3).

\clearpage
\newpage

\begin{figure}
	\centering
	\includegraphics[width=0.55\linewidth]{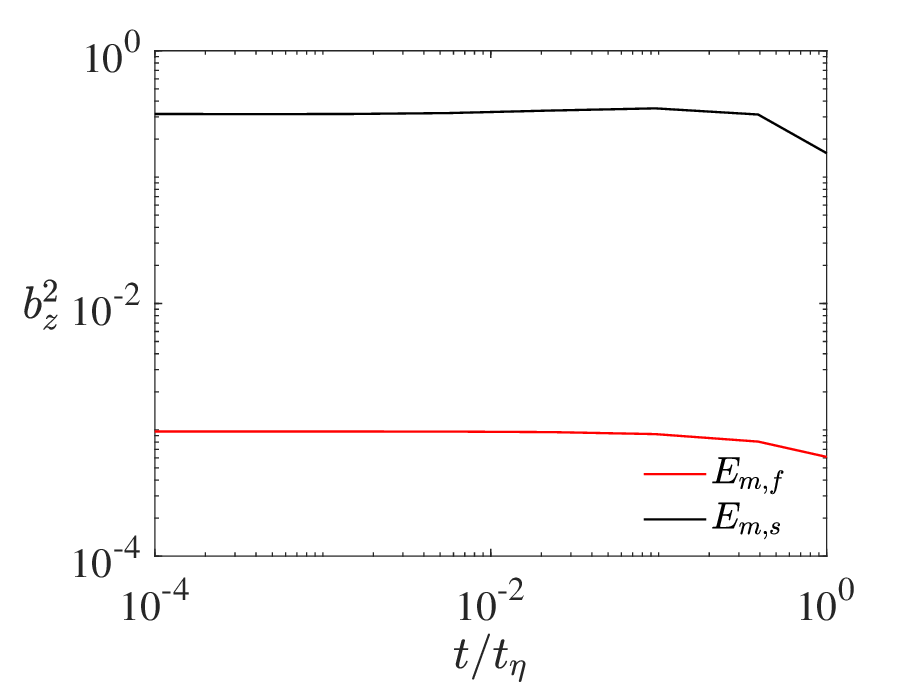}
	\caption{The total magnetic energy associated with fast
		($f$) and slow ($s$) MAC waves. 
		The parameters used are $E_\eta = 1 \times 10^{-5}$, 
		$B_0^2=500$ and $t/t_\eta = 7.2 \times 10^{-3}$.}
	\label{em}
\end{figure}

\clearpage
\newpage

\begin{figure}
	\centering
	\hspace{-2 in}	(a)  \hspace{2 in} (b)  \hspace{2 in} (c) \\
	\includegraphics[width=0.32\linewidth]{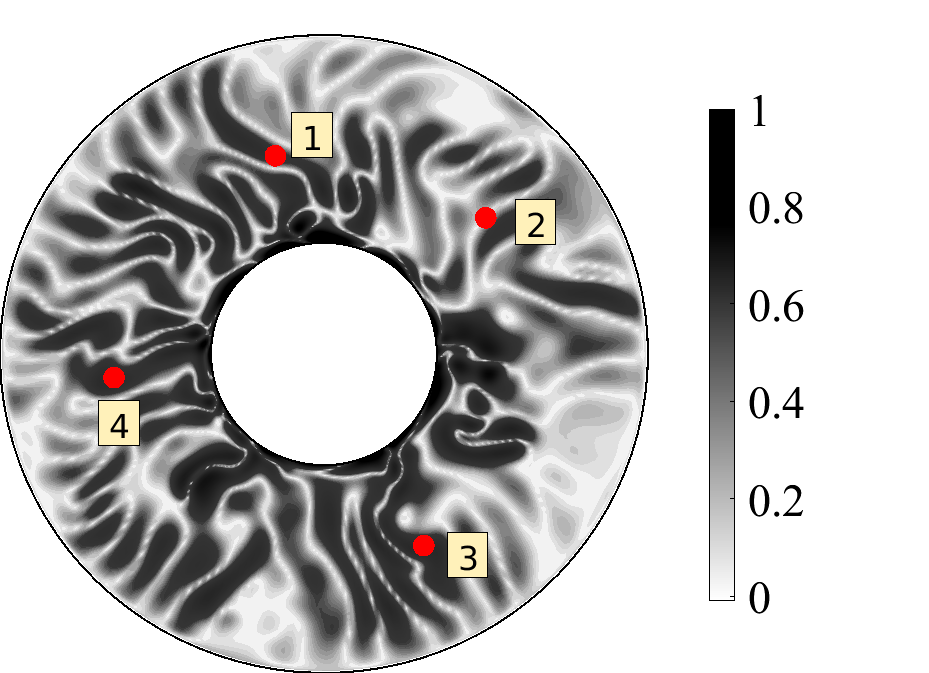}
	\includegraphics[width=0.32\linewidth]{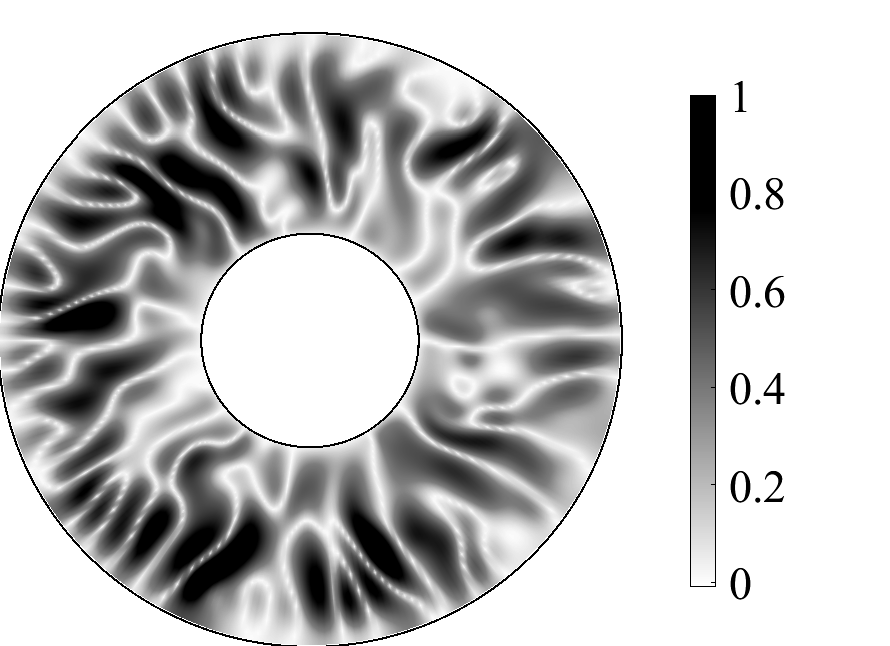}
	\raisebox{-0.4cm}
	{\includegraphics[width=0.32\linewidth]{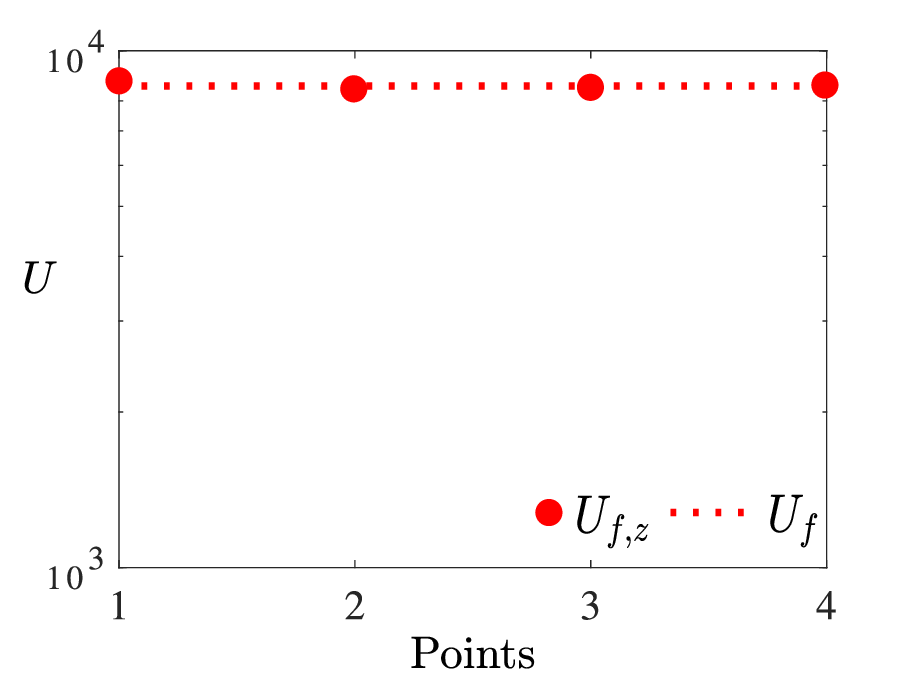}}
	\caption{Greyscale plots on a horizontal section 
		at $z = 0.2$ below the equatorial plane for 
		a hydrodynamic simulation, showing the ratio 
		of the magnitudes of (a) Coriolis and 
		(b) buoyancy force terms in the 
		$z$-vorticity equation to the magnitude of 
		the largest force among them. 
		(c) The estimated and measured group 
		velocities for the red points 
		in figure (a). Here, $U_{f,z}$ 
		is the measured axial 
		group velocitiy of the fast MAC waves and 
		$U_{f}$ is 
		their mean estimated axial group velocity.
		The other dynamo parameters are $Ra^C=3000,
		Ra^T=220, E = 6 \times 10^{-5}$, and 
		$Pm=Sc=5,Pr=0.5$. }
	\label{forceshydro}
\end{figure}

\clearpage
\newpage

\begingroup
\setlength{\tabcolsep}{12pt} % Default value: 6pt
\renewcommand{\arraystretch}{0.9} % Default value: 1
%\begin{landscape}
\begin{longtable}{ccccccccc}
	%	\hline
	\toprule
	$Ra^C$&$Ra^T$
	&$N_r$&$l_{max}$&$l_E$&$l_C$&$\bar{m}$&$\bar{k}_s$&$\bar{k}_z$\\
	\midrule
	\endfirsthead
	\toprule
	$Ra^C$&$Ra^T$
	&$N_r$&$l_{max}$&$l_E$&$l_C$&$\bar{m}$&$\bar{k}_s$&$\bar{k}_z$\\
	\midrule
	\endhead
	\bottomrule
	\multicolumn{9}{r}{{Continued on next page}} \\
	\endfoot
	\caption{The resolutions and calculated wavenumbers of the dynamo 
		simulations are presented. Here, $Ra^C$ and $Ra^T$ are the modified 
		compositional and thermal Rayleigh numbers respectively,
		$N_r$ is the number of radial grid points; 
		$l_{\text{max}}$ is the maximum spherical harmonic degree; $l_C$ 
		and $l_E$ are the mean spherical harmonic degrees 
		of convection and energy injection 
		respectively, $\bar{m}$ is the mean spherical harmonic order 
		in the range $l \leq l_E$, and $\bar{k}_s$ and $\bar{k}_z$ are 
		the mean wavenumbers along the cylindrical
		radius $s$ and axis $z$ in the range $l \leq l_E$.}
	\label{tableextra}\\
	%	\bottomrule
	%	\multicolumn{9}{r}{{End of the Table}} \\
	\endlastfoot
	\multicolumn{9}{c}{$E = 6 \times 10^{-5},Pm=5,Sc=5$}\\
	\multicolumn{9}{c}{ Pure composition}\\
	1000    &0&132&132&20&12&6&3.67&4.14\\
	2000    &0&132&132&20&13&6&3.55&3.99\\
	2500    &0&132&132&22&14&6&3.94&3.93\\
	2750    &0&132&132&24&14&7&3.52&4.00\\
	3000    &0&132&132&25&14&7&3.48&4.35\\
	4000    &0&132&132&25&14&7&3.11&3.53\\
	\multicolumn{9}{c}{ Pure thermal}\\
	%	0&220&96&96&	\multicolumn{5}{c}{ Onset thermal Rayleigh number}\\
	0&440&96&96&19&9&5&3.96&4.34\\
	\multicolumn{9}{c}{ Two-component convection}\\
	%	\multicolumn{11}{c}{ Thermal profile is purely internal heating}\\
	1000   &220    &132&132&16&11&7&4.48&4.81\\
	3000   &220   &132&132&17&12&7&4.22&4.53\\	
	6000   &220  &132&132&25&15&7&4.52&4.26\\	
	8000   &220  &176&176&26&15&8&4.42&4.53\\
	9000   &220  &176&176&26&15&8&4.31&4.31\\	
	12000  &220    &176&176&27&16&8&4.52&4.01\\	
	14000  &220    &176&176&28&16&8&3.82&3.91\\
	15000  &220    &192&192&28&16&8&3.79&3.83\\
	16000  &220    &192&192&29&16&8&3.79&3.88\\
	17000  &220    &192&192&29&16&8&3.65&3.43\\
	18000  &220  &192&192&29&16&8&3.82&3.53\\
	20000  &220    &192&192&30&16&8&3.71&3.45\\
	
	\multicolumn{9}{c}{ }\\
	%	3000&0     &132&132&25&14&7&5.12&4.35\\
	%	3000&50    &132&132&25&14&7&5.33&4.64\\
	%	3000&75    &132&132&25&14&7&5.36&4.17\\
	%	3000&100   &132&132&24&14&7&5.44&4.78\\
	3000&150   &132&132&21&13&7&4.59&4.11\\
	3000&220   &132&132&17&12&7&4.22&4.53\\	
	3000&440   &192&180&17&12&6&3.77&4.28\\
	3000&500   &176&176&19&12&6&4.08&4.25\\
	3000&660   &165&165&22&12&6&3.93&4.41\\
	3000&750   &198&192&22&12&6&4.17&4.53\\
	3000&1000  &176&176&22&12&6&3.68&4.48\\
	\multicolumn{9}{c}{ }\\
	9000&440&176&176&24&14&8&4.26&3.93\\	
	\multicolumn{9}{c}{ }\\
	\multicolumn{9}{c}{ $E = 1.2 \times 10^{-5},Pm=2,Sc=2$}\\
	\multicolumn{9}{c}{ Pure composition}\\
	1000    &0&144&144&31&19& 9&3.84&4.83\\
	3000    &0 &144&144&32&19& 9&5.56&3.96\\
	6000    &0  &192&192&35&20& 9&4.79&3.88\\
	9000    &0 &192&192&36&20&10.5&5.63&4.09\\
	10500   &0 &192&192&36&20&10&5.38&3.98\\
	12000   &0 &192&192&42&23&11.5&4.80&3.74\\
	\multicolumn{9}{c}{ Two-component convection} \\
	12000 &500&192&192&30&17&7&4.61&4.32\\ 
	16000 &500&192&192&32&18&7&4.96&4.16\\ 
	20000 &500&192&192&35&19&7&4.28&4.86\\ 
	
\end{longtable}
%\end{landscape}
\endgroup

\end{document}